\documentclass{article}

\usepackage{epsf}
\usepackage{longtable}
\usepackage{natbib}

\usepackage{graphicx}

\newcommand{\apj}{{Astrophys. J.}}
\newcommand{\mnras}{{Mon. Not. R. Astr. Soc.}}

\newcommand{\nat}{{Nature}}
\newcommand{\solphys}{{Solar Phys.}}
\newcommand{\apjl}{{Astrophys. J.}}
\newcommand{\aap}{{Astron. \& Astrophys.}}
\newcommand{\aaps}{{Astron. \& Astrophys. Suppl. Ser.}}
\newcommand{\apjs}{{Astrophys. J. Suppl.}}
\newcommand{\pasj}{{Proceedings of the Astronomical Society of Japan}}
\newcommand{\grl}{Geophys. Res. Lett.}
\newcommand{\jgr}{J. Geophys. Res.}

\begin{document}

\title{Solar Interior Rotation and its Variation}

\author{Rachel Howe\\
National Solar Observatory, \\
950 N.\ Cherry Ave.,\\
Tucson AZ 85719, U.S.A.\\
rhowe@noao.edu\\
http://www.noao.edu/staff/rhowe/}

\date{}
\maketitle

\begin{abstract}
This article surveys the development of observational understanding of
the interior rotation of the Sun and its temporal variation over
approximately forty years, starting with the 1960s attempts to
determine the solar core rotation from oblateness and proceeding
through the development of helioseismology to the detailed modern
picture of the internal rotation deduced from continuous helioseismic
observations during solar cycle~23. After introducing some basic
helioseismic concepts,  it covers, in turn, the rotation of the core
and radiative interior, the ``tachocline'' shear layer at the base of
the convection zone, the differential rotation in the convection zone,
the near-surface shear, the pattern of migrating zonal flows known as
the torsional oscillation, and the possible temporal variations at the
bottom of the convection zone. For each area, the article also briefly
explores the relationship between observations and models.
\end{abstract}


\newpage


\section{Introduction}
\label{section:introduction}

The internal rotation of the Sun is intimately related to the
processes that drive the activity cycle. \citet{1989ApJ...343..526B} stated that, ``Knowledge of the
internal rotation of the Sun with latitude, radius, and time is
essential for a complete understanding of the evolution and the
present properties of the Sun,'' and this remains true
today. 

The Sun rotates on its axis approximately once every twenty-seven
days; however, the rotation is not uniform, being substantially slower
near the poles than at the equator. This superficial 
aspect of the solar differential rotation was well known from 
sunspot observations as early as the seventeenth century.
However it is only within the
last thirty years that it has become possible to observe the rotation
profile in the solar interior, and mostly within the most recent solar
cycle that its subtle temporal variations have become
evident.
Helioseismology -- the study of the waves that propagate within
the Sun and the inference from their properties of the solar interior
structure and dynamics -- is the most important tool we have to measure
this internal rotation.

In this review, we start by introducing some of the basic concepts of
helioseismology (\S~\ref{section:acoustic}) and the inversion problem 
(\S~\ref{section:inversionbasics}) as it applies to the 
internal solar rotation. Next, after a brief historical overview (\S~\ref{section:obs}) of the observations, we consider what we have learned from
helioseismology about the rotation profile and its variation with
depth. 

\begin{figure}[htbp]
\centerline{\includegraphics[width=5.5in]{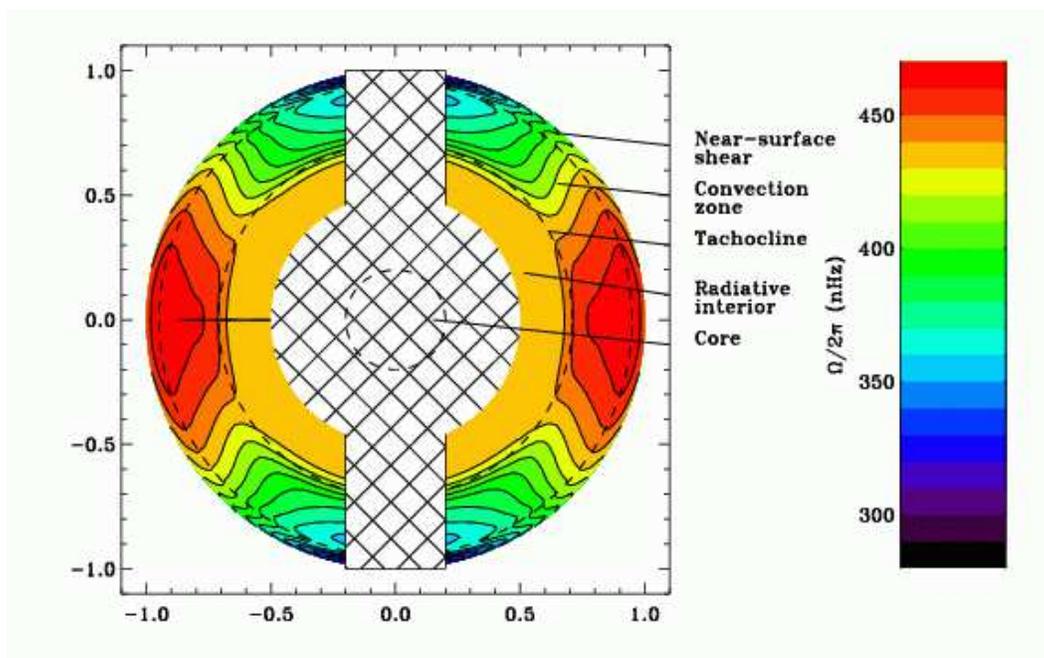}}
\caption{A section through the interior of the Sun,
showing the contours of constant rotation and the major features of the 
rotation profile, for a temporal average over about twelve years of 
MDI data. The 
cross-hatched areas indicate the regions in which it is difficult 
or impossible to obtain reliable inversion results with the available
data.}
\label{fig:quad}
\end{figure}

We consider first the time-invariant part of the solar rotation
profile. The main features of interest are (Figure~\ref{fig:quad}):
\begin{enumerate}
  \item the radiative interior and core, which 
appear to rotate approximately as a solid body,
though the innermost core may behave differently;
 (\S~\ref{section:interior}),
  \item the tachocline, a relatively thin zone of shear between the differentially rotating convection zone and the radiative interior, which is believed to
play an important role in the solar dynamo
(\S~\ref{section:tachocline});
  \item the differential rotation in the bulk of the convection zone (\S~\ref{section:bulk}); and 
  \item the subsurface shear layer between the fastest-rotating
layer at about $0.95R_\odot$ and the surface. (\S~\ref{section:nearsurf}). 
\end{enumerate}

We will consider each of these in turn, working outwards from the core
to the surface, and then discuss the time-varying part of the rotation
-- the torsional oscillation (\S~\ref{section:torsional}) and the possible variations at the base
of the convection zone (\S~\ref{section:tachvar}). We attempt to place the observations in the
context of models; however, this is a review from an observer's point
of view, and an exhaustive examination of the models themselves is
beyond its scope.

\newpage


\section{Acoustic Modes}
\label{section:acoustic}

\begin{figure}
\centerline{\includegraphics[width=5in]{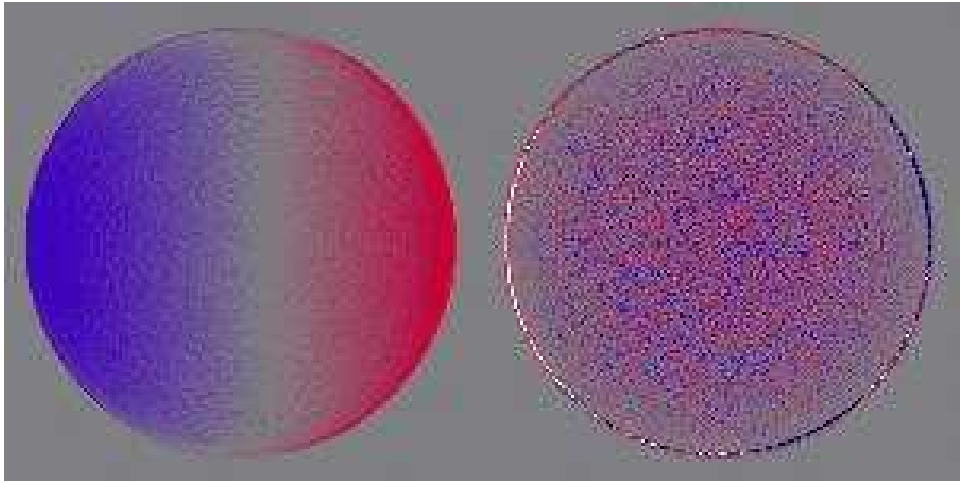}}
\caption{\label{fig:doppler}A single Doppler velocity image of the
Sun from one GONG [Global Oscillation Network Group] instrument (left),
and the difference between that image and one taken a minute earlier (right).
with red corresponding to motion away from, and blue to motion towards, the 
observer.
The shading across the first image comes from the solar rotation.}
\end{figure}

\begin{figure}
\centerline{\includegraphics[width=5in]{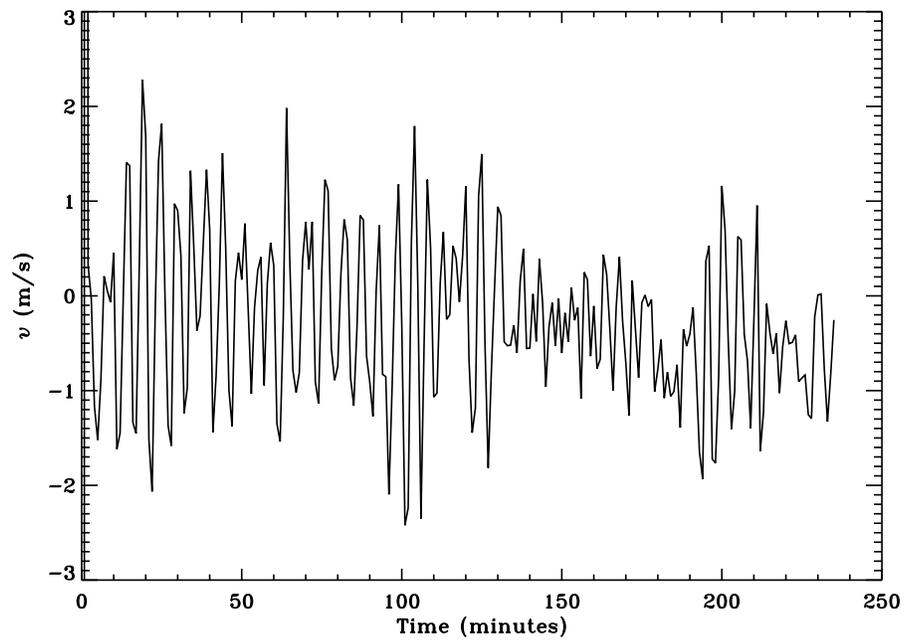}}
\caption{\label{fig:tseries} A segment of an $l=0$ time
series of Doppler velocity observations, showing the
dominant five-minute period and the rich beat structure.
}
\end{figure}

The raw data of helioseismology consist of measurements of the photospheric
Doppler velocity -- or in some cases intensity in a particular wavelength band -- taken  at a cadence of about one minute and generally collected with as little interruption as possible over periods of months or years; the measurements can be either imaged or integrated (``Sun as a Star"). An overview of the
observation techniques can be found in \citet{1991soia.book..329H}.
Figure~\ref{fig:doppler} shows a typical single Doppler velocity image
of the Sun, and Figure~\ref{fig:tseries} 
a portion of an $l=0$ time series, derived by averaging the 
velocity over the visible disk for each successive image in a
set of observations. The five-minute period and the
rich beat structure are clearly visible in the time series. For
an example of an integrated-sunlight spectrum from a 
long series of observations, see Figure~\ref{fig:bspec}. 

\begin{figure}[!hbpt]
\centerline{\includegraphics[width=4in]{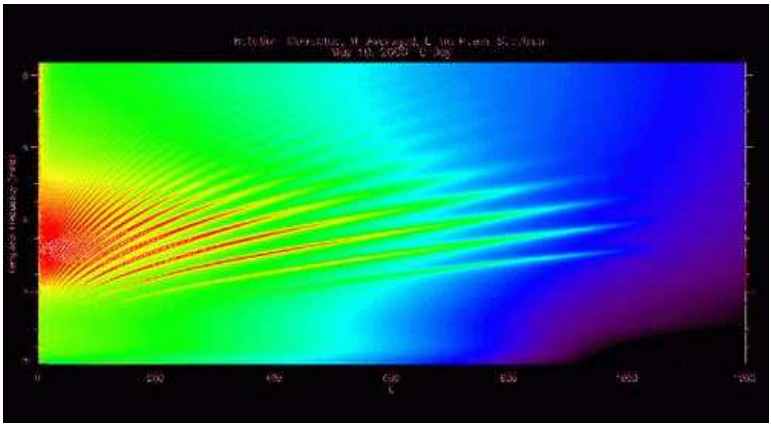}}
\caption{\label{fig:lnu} Typical $l-\nu$ spectrum from one day of 
GONG observations. (Image courtesy NSO/GONG.)}
\end{figure}

As was first discovered by
\citet{1975A+A....44..371D}, the velocity or intensity variations at the solar surface have a 
spectrum in $k-\omega$ or $l-\nu$ space that reveals their origin in 
acoustic modes propagating in a cavity bounded above by the solar surface
and below by the wavelength-dependent depth at which the waves are refracted
back towards the surface. These ``$p$ modes" can be classified by their
radial order $n$, spherical harmonic degree $l$, and azimuthal order $m$;
as discussed, for example, in \S~2.2 of \citet{lrsp-2005-6},
the radial displacement of a fluid element 
at time $t$, latitude $\theta$ and longitude $\phi$ can be written in the form
\begin{equation}
\delta r(r,\theta,\phi,t)=\sum_{m=-l}^l{a_{nlm}\xi_nl(r)Y_l^m(\theta,\phi) e^{i\omega_{nlm}t}},
\end{equation} 
where $\xi_{nlm}$ is the radial eigenfunction of the mode with
frequency $\omega_{nlm}$ and $Y_l^m{\theta,\phi}$ is a spherical 
harmonic.
As seen in Figure~\ref{fig:lnu}, the power in the spectrum
falls along distinct ``ridges'' in the $l-\nu$ plane, each
ridge corresponding to one radial order. The modes making up the
$n=0$ ridge are the so-called $f$ modes, which are surface gravity 
waves. The $p$  
modes, so called because their restoring force is pressure,  are excited at the surface and have their largest amplitudes
there. Another class of modes, the $g$ modes with gravity
as the restoring force, excited in the core
and with amplitudes vanishing at the surface, are hypothesized to exist
but have so far not been definitely observed (\S~\ref{section:gmodes}).

\begin{figure}
\centerline{\includegraphics[width=4in]{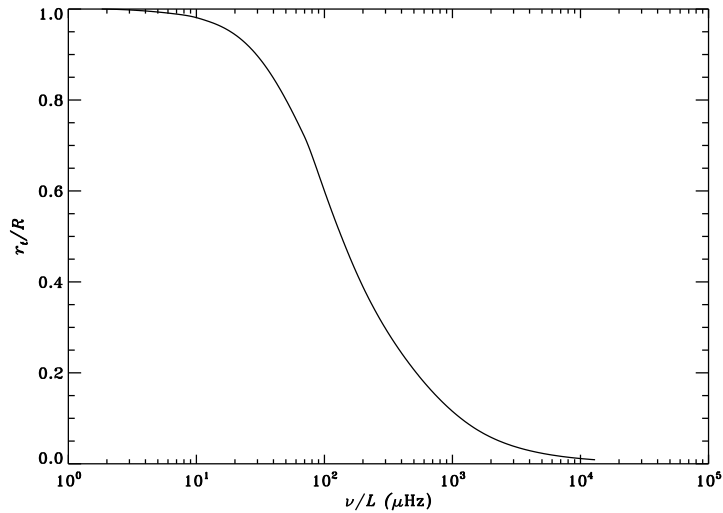}}
\caption{\label{fig:wxt}Lower turning point of 
acoustic modes as a function of phase speed $\nu/L$,
based on Model S of \citet{1996Sci...272.1286C}.}
\end{figure}

\begin{figure}[htbp]
\centerline{\includegraphics[width=5.5in]{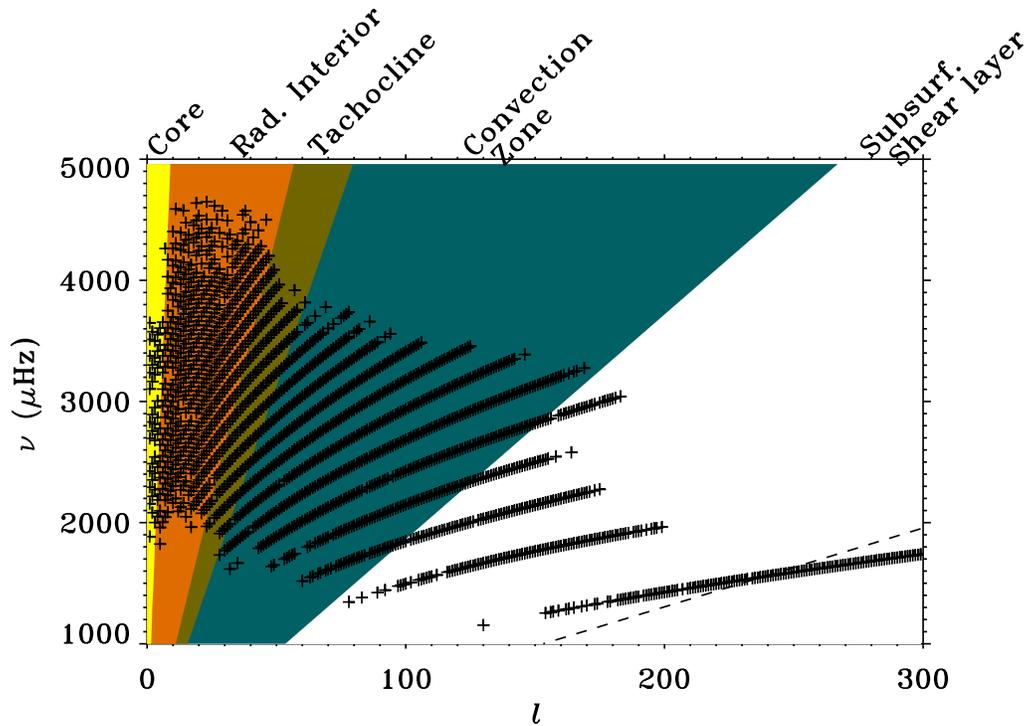}}
\caption{Locations
of modes in the $l,\nu$ plane for a typical
MDI mode set. The colored shading represents the radial regions in which the
modes have their lower turning points; 
the core, $r\leq 0.2R_\odot$, 
the radiative interior, $0.2 \leq r/R_\odot \leq 0.65$, the 
tachocline, $0.65 \leq r/R_\odot \leq 0.75$,
the bulk of the convection zone,  $0.75 \leq r/R_\odot \leq 0.95$,
and the near-surface shear layer, $r/R_\odot \geq 0.95$; 
the dashed line
on the lower right corresponds to $r/R_\odot = 0.99$. 
}
\label{fig:lnufig}
\end{figure}

The longer the horizontal wavelength -- and the lower the degree -- the more
deeply the modes penetrate, with the radial $l=0$ mode going all the
way to the core of the Sun (but providing no rotational information),
while modes with $l\ge 200$ or so penetrate only a few megameters below the
surface and are generally too short-lived to form global standing waves; 
these are the modes used for local helioseismology. The lower turning 
point radius, $r_t$, is a monotonic function of the
phase speed $\nu/L$, where $L =\sqrt{l(l+1)} \approx l+1/2$,
as shown in Figure~\ref{fig:wxt}.
The varying penetration depth with degree, as illustrated in Figure~\ref{fig:lnufig}, makes it possible to 
deduce the rotation and other properties of the
solar interior profile as a function of depth.

The lowest-degree modes are observed in integrated sunlight,  
but for the purposes of 
measuring the interior rotation profile we are mostly concerned with 
what are termed medium-degree ($l \leq 300$) modes, which can be observed
with imaging instruments of relatively modest ($\approx$~10~arcsec) resolution.
The power in the modes peaks at about 3~mHz, or a period of 5~minutes;
useful measurements can be made for modes between about 1.5 and 5~mHz, with the
frequency determination becoming more challenging at the extremes 
due to signal-to-noise issues and, at the high-frequency end, to the
increasing breadth of the peaks.

\subsection{Differential rotation and rotational splitting}
\label{section:diffsplit}

The Sun's rotation lifts the degeneracy between modes of the same $l$ and different $m$, resulting in ``rotational splitting'' of the frequencies as
waves propagating with and against the direction of rotation
(prograde and retrograde) have higher and lower frequencies. To 
first order, the splitting $\delta\nu_{m,l}\equiv  \nu_{-m,l}-\nu_{+m,l}$ 
is proportional to the rotation rate multiplied by $m$.
 
\begin{figure}
\centerline{\includegraphics[width=5.0in]{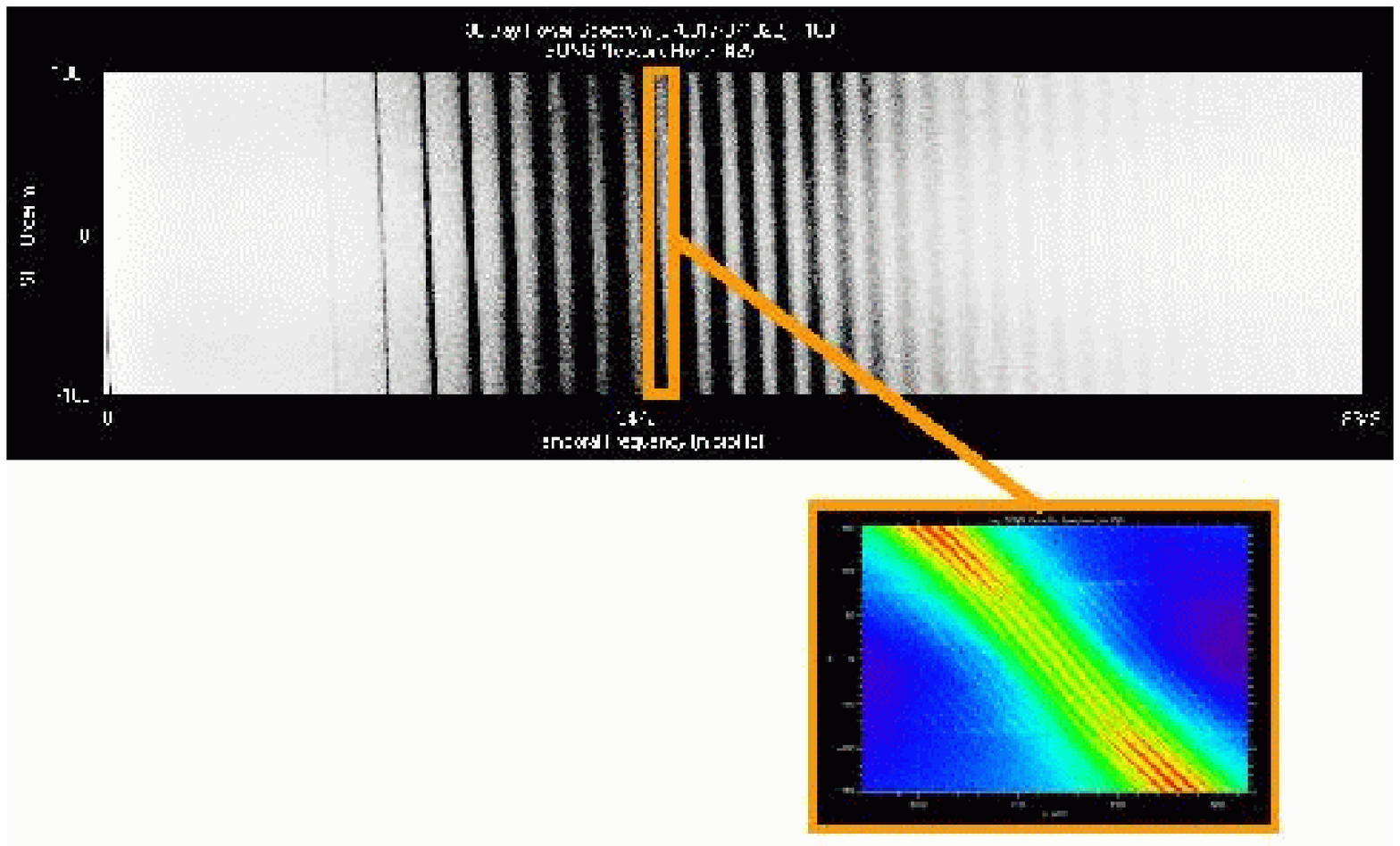}}
\caption{\label{fig:mnu}Spectrum for $l=100$ in the $\nu,m$ 
plane (top) and detail (bottom) of a single ridge (radial order)
showing the curvature due to differential rotation and the
multiple-ridge structure arising from spherical harmonic
leakage. }
\end{figure}

Figure~\ref{fig:mnu} shows a typical $m-\nu$ acoustic spectrum of GONG data at
$l=100$. The effect of the rotation causes the ridges at each $n$ to 
slant away from the $\nu=0$ axis; closer examination reveals that the
ridges have an S-curve shape that arises from the differential rotation, 
and also shows the ridge structure due to leakage, which will be
discussed below in \S~\ref{section:leaks}.

\begin{figure}[htbp]
\centerline{\includegraphics[width=5.5in]{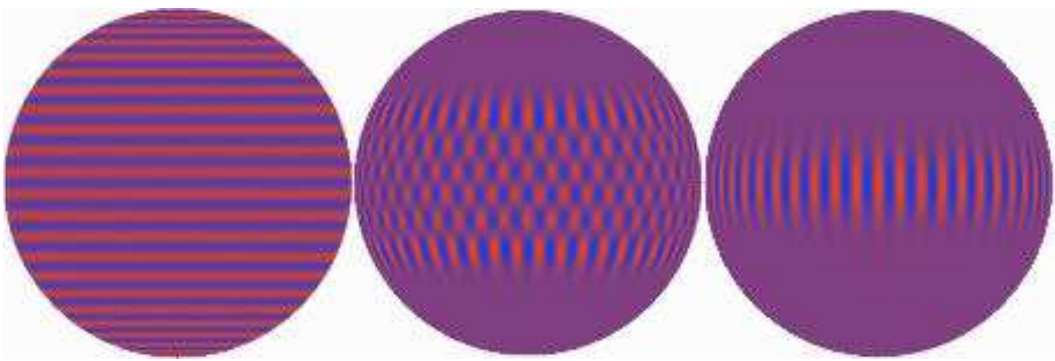}}
\caption{Spherical harmonic patterns for 
$l=50$: (left, $m=0$; center, $m=45$, right; $m=50$).}
\label{fig:harmfig}
\end{figure}

Because 
modes of different $m$ values sample different latitude ranges, with the
sectoral ($|m|=l$) modes confined to a belt around the
equator and the zonal or $m=0$ modes reaching to the poles, 
as illustrated in Figure~\ref{fig:harmfig}, we can measure the
rotation as a function of latitude.


A given ($n,l$) multiplet consists of $2l+1$ frequency measurements if
each ($l,m$) spectrum is analyzed separately, 
though some fraction 
of these frequencies may be missing in any given data set. This amount of 
data was somewhat unwieldy in the early days of helioseismology. It is
therefore common to express $\nu_{nl}(m)$ as a polynomial expansion,
for example,
\begin{equation}
\nu_{nlm}
= \nu_{nl} + \sum_{j=1}^{j_{\rm max}} a_j (n,l) \, {\cal P}_j^{(l)}(m), 
\label{eq:eq0}
\end{equation}
where the basis functions are polynomials related
to the Clebsch-Gordan coefficients 
$C_{j0lm}^{lm}$ by
\begin{equation}
{\cal P}_j^{(l)}(m) = {l\sqrt{(2l-j)!(2l+j+1)!}\over
(2l)!\sqrt{2l+1}}C_{j0lm}^{lm}
\label{eq:eq0b}
\end{equation}
\citep{1991ApJ...369..557R}.
Indeed,
in many analysis schemes
coefficients of the expansion are derived by fitting directly to the
acoustic spectrum and the individual frequencies are not measured. This
approach can improve the stability of the fits, perhaps at the cost
of imposing systematic errors. 
Early work used Legendre polynomials; however, most modern work
uses either Clebsch--Gordan coefficients or the Ritzwoller--Lavely
formulation, 
which come closer to being truly orthogonal 
for the solar rotation problem.
Only the odd-order coefficients encode the rotational asymmetry, while the
even-order coefficients contain information about the structural 
asphericity. Roughly speaking, the $a_1$ coefficient 
describes the rotation rate averaged over all latitudes,
and the $a_3$ and higher coefficients describe the differential rotation.

\subsection{Spherical Harmonics and Leakage}
\label{section:leaks}

Spherical harmonic masks are used to separate the contributions from 
modes of different degree and azimuthal order into complex time series,
which can then be transformed to acoustic Fourier spectra.

The radial component of the velocity at the solar surface from a 
mode with a given degree $l$, azimuthal order $m$ and radial 
order $n$ is given by 
\begin{equation}\label{eq:v}
V_{n,l,m}(\phi,\theta,t)={\rm Re}[a_{n,l,m}(t)P_l^{|m|}(\cos{\theta})e^{im\phi})],
\end{equation}
where Re[] denotes the real part, $\phi$ is longitude and $\theta$ is
latitude. (See, for example, \citealt{1994A+AS..107..541S}.)
The masks used separate the different spherical harmonics
take the form
\begin{equation}
M_{l,m} \propto {Y_{l,m}(\theta,\phi)A(\rho)},
\end{equation}
where $A$ is an apodization function and $\rho\equiv \sqrt{\cos^2\theta+\sin^2\theta\sin^2\phi}$ represents the 
fractional distance
from disk center in the solar image. The line-of-sight projection factor
is $V=\sqrt{1-\rho^2}$. 

Because only part of the
solar surface is visible at any time, the masks are not completely 
orthogonal and the modes ``leak'' into neighboring spectra. This leakage
complicates the analysis and can cause systematic errors in the
measured frequencies if it is not correctly taken into account. For 
a detailed discussion of the calculation of the so-called ``leakage matrix,'' 
see \citet{1994A+AS..107..541S,1998ESASP.418..225H}. Briefly, 
the leakage matrix element $s(l,m,l^\prime,m^\prime)$ for leakage from the 
$l^\prime,m^\prime$ mode to the $l,m$ spectrum can be computed as
\begin{equation}
s(l,m,l^\prime,m^\prime)={1\over\pi}\int_{-1}^{1}\int_{-\pi/2}^{\pi/2}
{P_l^m(x)P_{l^\prime}^{m^\prime}(x) \cos(m\phi)\cos(m^\prime \phi)
V(\rho)A(\rho)dxd\phi}.
\end{equation}
Symmetry properties in this expression lead to some simple exclusion 
rules; for example, leaks with odd $|\delta l+\delta m|$ (where
$\delta m \equiv m-m^\prime$ and $\delta l \equiv l-l^\prime$)
are not allowed. 


One example of the importance of the leakage is in the contribution of the  
so-called $m$-leaks ($\delta l=0,\delta m =\pm 2$) to the estimation
of low-degree splittings.
As pointed out, for example, by \citet{1998AAS..131..539H}, these leaks
are strongest for small $|m|$; they are also asymmetrical, 
especially for $|m|=l$, where the $m=l$ peak has an $m=l-2$ leak
on one side and no counterbalancing $m=l+2$ leak on the other. Especially
for $l=1$, this can introduce a serious systematic error into the estimate
of the splitting if not properly accounted for. 

Leakage also means that integrated-sunlight 
instruments (which effectively use the $l=0$ mask) 
can detect modes of $0 \leq l \leq 5$, though the 
sensitivity falls off rapidly for $l>1$. All these modes appear in 
a single acoustic spectrum; the instruments are sensitive to odd-$m$ modes for 
odd $l$ and to even-$m$ modes for even $l$, with the sectoral, or $|m|=l$,
modes most strongly detected.  

In general, the leakage has effects throughout the acoustic spectrum, but the 
most deleterious effects arise when the leaks cannot be resolved from the
target peaks. This occurs for $m$-leaks at frequencies above about 2~mHz;
for higher-degree modes the leakage between modes of adjacent $l$ 
becomes a problem, as the ridges become both broader, and more closely spaced in frequency, at around $l=150$. Beyond this point the peaks cannot be fitted independently, and some modeling of the leakage is essential in order
to estimate the mode parameters.


\begin{figure}[htbp]
\center\includegraphics[width=4in, angle=90]{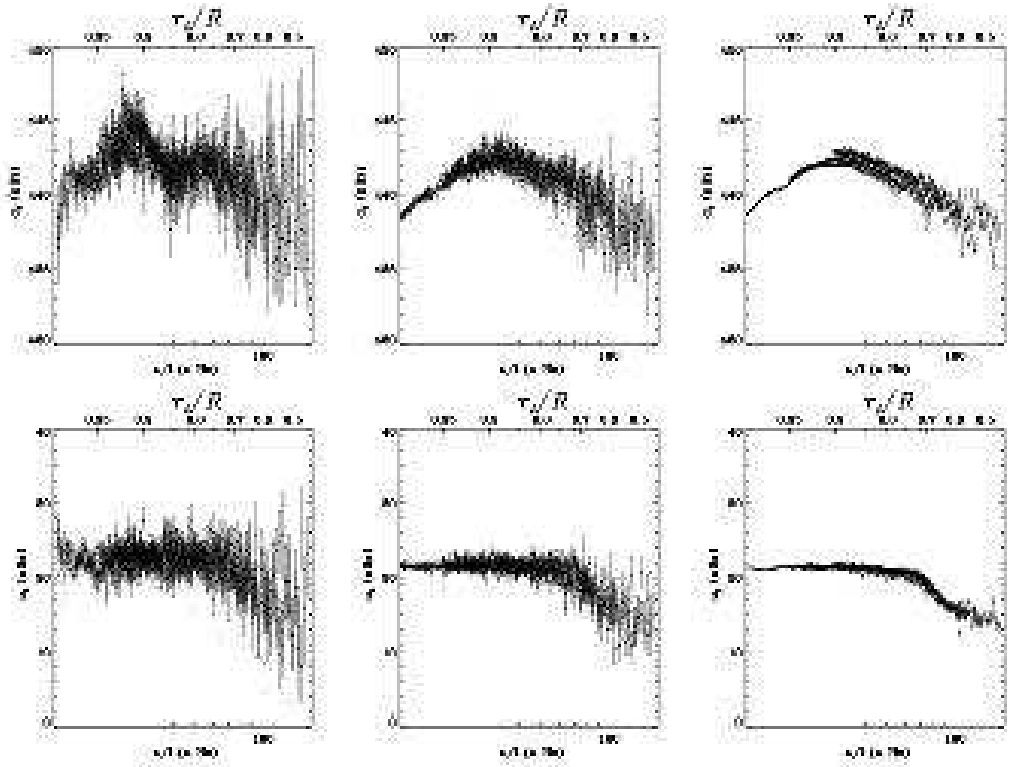}
\caption{$a_1$ (top) and $a_3$ (bottom) 
coefficients for (left), 1986 BBSO observations, (middle) 108 days of
GONG observations in 1996, (right) the mean of 35 consecutive 108-day periods
of GONG observations from 1995\,--\,2005, plotted as a function of 
phase speed with the turning point radius marked on the upper axis.}
\label{fig:coeffs}
\end{figure}

\subsection{Estimating rotation properties directly from coefficients}
\label{section:coeffcomb}

It is possible to make some inferences about the rotation profile without carrying out a full-scale inversion.
Simple examination of the odd-order coefficients, sorted by
the lower turning-point radius of the modes, reveals the existence of the
near-surface shear, the differential rotation within the convection zone,
and a discontinuity in the differential rotation at the base of the
convection zone, as shown in Figure~\ref{fig:coeffs}. More sophisticated analysis is also possible. For example, \citet{1995ApJ...438..445W} gave approximate expressions for the rotation profile at different 
latitudes as sensed by a particular $n,l$ multiplet, $\bar\Omega^{nl}$, as
follows:
\begin{eqnarray}
  \bar\Omega_0^{nl} \approx a_1^{nl}+a_3^{nl}+a_5^{nl},\\
  \bar\Omega_{30}^{nl} \approx a_1^{nl}-{{a_3^{nl}}\over 4} - {{19a_5^{nl}} \over 16}, \\
  \bar\Omega_{45}^{nl} \approx a_1^{nl}-{{3a_3^{nl}}\over 2} - {{3a_5^{nl}} \over 4}, \\
  \bar\Omega_{60}^{nl} \approx a_1^{nl}-{{11a_3^{nl}}\over 4} + {{37a_5^{nl}} \over 16}. 
  \label{eq:eqpw2}
\end{eqnarray}
These estimates, where the subscripts on the LHS refer to the
latitude in degrees, are noisy for individual multiplets, but \citet{1995ApJ...438..445W} were able to build up a picture of the internal rotation from BBSO data
by forming cumulative averages with the input data sorted in ascending
order of $\nu/L$.

\newpage


\section{Inversion Basics}
\label{section:inversionbasics}

Various inversion techniques are used to infer the internal rotation
profile from the observed frequency splittings. The aim of the inversion
procedure is to form linear combinations of the data that give
well-localized inferences of the rotation at a particular location within
the Sun. We will discuss only linear inversion methods,  as non-linear approaches are not needed for the relatively low velocities involved in the global rotation

\subsection{The inversion problem}
\label{section:inversionprob}

\begin{figure}[htbp]
\center\includegraphics[width=4in]{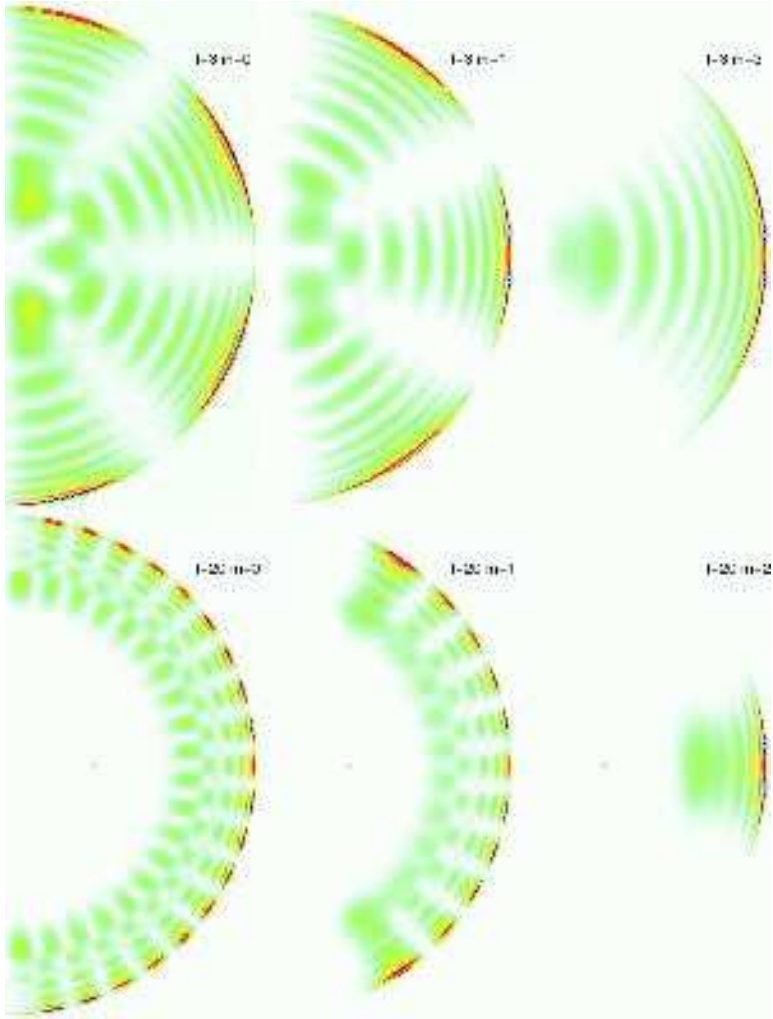}
\caption{Sections through rotation kernels for selected
azimuthal orders for $l=3,n=9$ (top) and $l=20,n=5$ (bottom).
\label{fig:watermelon}}
\end{figure}

The basic 2-dimensional rotation inversion problem can be stated as follows: we have a number $M$
of observations $d_i$, from which we wish to infer the
rotation profile $\Omega(r,\theta)$ where $r$ is distance from the
center of the Sun, 
and 
$\theta$ is (conventionally) colatitude. Each datum is a spatially
weighted average of the rotation rate:
\begin {equation}\label{eq:eq1}
d_i=\int_0^{R_\odot}\int_0^\pi{K_i(r,\theta)\Omega(r,\theta) dr d\theta} +\epsilon_i,
\end{equation}
where $R_\odot$ is the solar radius,
the error term $\epsilon$
corresponds to the noise and measurement error in the data, and $K$ is a model-dependent spatial weighting function known as the {\em kernel} \citep{1977ApJ...217..151H,1980A+A....89..207C}. For the two-dimensional rotation inversion, the radial part 
is related to the eigenfunction of the mode and the latitudinal part to
the associated Legendre polynomial; \citet{1994ApJ...433..389S} give the
expression for the kernel as 
\begin{equation}\label{eq:eq1b}
K_{nlm}(r,\theta)= {m\over {I_{nl}}}\left\{{{\xi_{nl}(r)\left[{\xi_{nl}(r)-{2\over L}\eta_{nl}(r)}\right]{P_l^m(x)^2}}}\right.
\end{equation}
\begin{eqnarray*}
+{{\eta_{nl}(r)^2}\over{L^2}}\left[\left({dP_l^m} \over{dx}\right)^2(1-x^2)
-  \left .{2P_l^m{{dP_l^m}\over{dx}}x} + {{m^2}\over{1-x^2}}P_l^m(x)^2\right]\right\}\rho(r)r\sin{\theta},
\end{eqnarray*}
where
\begin{equation}
I_{nl}=\int_0^{R_\odot}{[\xi_{nl}(r)^2+\eta_{nl}(r)^2]\rho(r)r^2 dr},
\end{equation}
$x=\cos{\theta}$, $L^2=l(l+1)$, $\xi_{nl}$ is the radial displacement 
for the eigenfunction of the mode, $L^{-1}\eta_{nl}$ is the 
horizontal displacement, and $\rho(r)$ is the density.
(See Figure~\ref{fig:watermelon} for illustrations of sample kernels.)

The aim of the inversion is to find
\begin{equation}\label{eq:eq2}
 \bar{\Omega}(r_0, \theta_0)=\sum_{i=1}^M{c_i(r_0,\theta_0)d_i},
\end{equation}
where $(r_0,\theta_0)$ is the location at which the
inferred rotation rate $\bar{\Omega}$ is to be found and the
$c_i$ are the coefficients to be used to weight the data;
the inversion process can be thought of as the search for the best values
for these coefficients.

\subsection{Averaging kernels}
\label{section:avkers}

By substituting Equation~(\ref{eq:eq1}) into the RHS of
Equation~(\ref{eq:eq2}) we obtain 
\begin{equation}\label{eq:eq3}
\bar{\Omega}(r_0,\theta_0) = \int_0^{R_\odot}\int_0^\pi{{\cal K}(r_0,\theta_0;
r,\theta)\Omega(r,\theta) dr d\theta} +\epsilon_i,
\end{equation}
where 
\begin{equation}\label{eq:eq4}
{\cal K}(r_0,\theta_0;r,\theta) \equiv \sum_{i=1}^M{c_i(r_0,\theta_0) K_i(r,\theta)}
\end{equation}
is the {\em averaging kernel} for the location $(r_0,\theta_0)$.
The averaging kernels are independent of the values of the data.
However, because the uncertainties in the data are used to weight the inversion calculation that generates the coefficients $c_i$, as described below in \S\S~\ref{section:rls} and \ref{section:ola}, these do enter indirectly into the averaging kernels. The averaging kernels also
depend on which modes are present in the input data set. They provide a useful 
tool for assessing the reliability of an inversion inference from
a particular mode set \citep[see, for example, ][]{1992ApJ...385L..59S,1994ApJ...433..389S}.

\subsection{Inversion errors}
\label{section:inverrs}

If the errors on the input data are uncorrelated and 
properly described by a normal distribution whose width
corresponds to the quoted uncertainty $\sigma_i$,
the formal uncertainty on the inferred profile is given by 
\begin{equation}
\sigma^2[\Omega(r_0,\theta_0)]=\sum_i[c_i(r_0,\theta_0)\sigma_i]^2.
\label{eq:erreq}
\end{equation}
In the (usually unrealistic) case where the errors on the input data
are all equal, we can write
\begin{equation}
\sigma^2[\Omega(r_0,\theta_0)]=\Lambda(r_0,\theta_0)\sigma,
\end{equation}
where the ``error magnification''  is
given by 
\begin{equation}
\Lambda(r_0,\theta_0)=\sum_i[c_i(r_0,\theta_0)^2]^{1/2}.
\end{equation}
As discussed, for example, by \citet{1990MNRAS.242..353C}, a
quantitative choice of regularization parameters can then be made
by finding the ``knee'' of a tradeoff curve where the error magnification
is plotted against the width of the averaging kernel. However, 
in the two-dimensional case this does not always give a clear result, and
this formulation of the error magnification is not very 
useful for modern data sets where the the uncertainties on the parameters
are anything but uniform. Instead, one can consider the uncertainty on the
inferred quantity at a particular location.

Even when the errors on the input data are uncorrelated,
the errors on the inferred profile will not be,
as discussed by \citet{1996MNRAS.281.1385H}. (As a simple way to understand
this, consider the case where one measurement is significantly ``off''; this will affect the inferred profile at every location where the inversion coefficient
$c_i$ for that datum is non-zero.)
In the one-dimensional case, the correlation between the errors
for two points $r_0$ and $r_1$ is given by 
\begin{equation}
C(r_0,r_1) = {{\sum{c_i(r_0)c_i(r_1)\sigma_i^2}}\over
{[\sum{c_i^2(r_0)\sigma_i^2}]^{1/2}[\sum{c_i^2(r_1)\sigma^2_i]^{1/2}}}};
\end{equation}
this can easily be generalized to the two-dimensional case. 
\citet{1996MNRAS.281.1385H} found that the spatial scale over which the
inversion errors are significantly correlated is usually similar to that for
the averaging kernels, though for some cases where the
inversion parameters have been badly chosen the results can be correlated over long distances even when the averaging kernels appear well formed.

Error correlations by definition should not distort the inferred profile
beyond the distribution predicted by the formal uncertainty on the inferences, 
provided always that the input uncertainties are correct. However, the finite width of
averaging kernels also gives rise to a systematic error that can be 
much larger. Consider, for example, the case where a thin shear layer 
is not resolved; then
all the estimated rotation rates on one side of the shear 
could be
underestimated, and those on the other side overestimated, by several times the formal uncertainty. Such systematic errors and their relationship to the 
averaging kernels have been discussed, for example, by \citet{1990MNRAS.242..353C}.

\citet{1996ApJ...459..779G} pointed out that it is not sufficient for 
the rotation rates at two locations to have 
non-overlapping errors as calculated in Equation~(\ref{eq:erreq}),
and 
described a method for increasing the
error estimates on inversions to allow truly significant differences 
between the inferred rotation rate at different locations to be determined.
This method, however, has not been widely used.

Because the input data are noisy and of finite resolution, the inversion 
problem does not have a unique solution; there will always be
a tradeoff between noise and good localization. Two widely-used 
approaches to balancing these criteria are ``regularized least squares" (RLS) and ``optimally localized
averaging'' (OLA).

\subsection{Regularized least squares}
\label{section:rls}

\begin{figure}[htbp]
\center\includegraphics[width=5in]{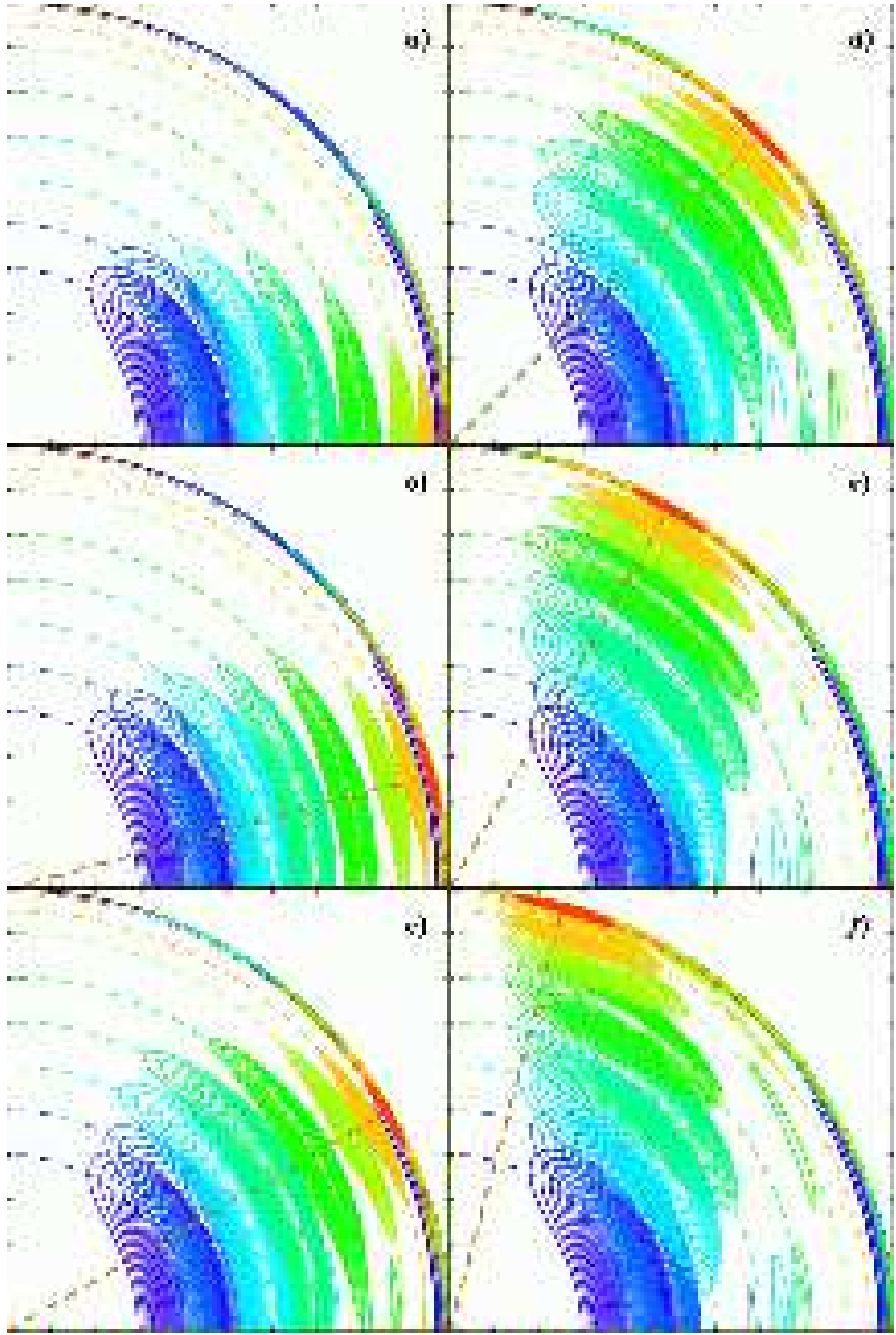}
\caption{Averaging kernels for a typical RLS inversion of MDI data,
  for target latitudes 0 (a), 15 (b), 30 (c), 45 (d), 60 (e) and 75
  (f) degrees as marked by the dashed radial lines, and target radii
  0.4, 0.5, 0.6, 0.7, 0.8, 0.9, 0.95, 0.99~$R_\odot$ indicated by colors
  from blue to red as denoted by the dashed concentric
  circles. Contour intervals are 5\% of the local maximum value
  closest to the target location, with dashed contours indicating
  negative values.}
\label{fig:ker1}
\end{figure}

The RLS approach to the inversion problem is to find (essentially
through a least-squares fit) the model profile that best fits the
data, subject to a smoothness penalty term, or regularization. 
More regularization -- a
larger weighting for the penalty term -- results in poorer spatial
resolution (and potentially more systematic error) but smaller uncertainties. In one such implementation \citep{1994ApJ...433..389S},
we minimize
\begin{equation}\label{eq:eq5}
\sum_i{{[{d_i-\int_0^R \int_0^{\pi}{\bar\Omega(r,\theta)K_i(r,\theta) dr d\theta]}^2}
\over{(\sigma_i/\bar{\sigma}})^2}} 
+ \mu_r^2\int_0^R\int_0^{\pi}{({{d^2\bar\Omega}\over{dr^2}})^2 dr d\theta}
+ \mu_\theta^2\int_0^R\int_0^{\pi}{({{d^2\bar\Omega}\over{d\theta^2}})^2 dr d\theta}
\end{equation}
with $\mu_r$ and $\mu_\theta$ being the radial and latitudinal tradeoff
parameters.
The RLS inversion has the advantages of being computationally
inexpensive and always (thanks to the 
second-derivative regularization, which amounts to
an a priori assumption of smoothness) providing some kind of estimate
of the quantity of interest
even in locations that are not, strictly speaking, resolved by the data.
In this method, the averaging kernels ${\cal K}$ can (but need not be)
calculated from the coefficients in a separate step. They are not
guaranteed to be well localized, though they are forced to have a center
of mass at the specified location $r_0,\theta_0$.
Figure~\ref{fig:ker1} illustrates typical averaging kernels
for a 2dRLS inversion of an MDI data set.

\begin{figure}[htbp]
\center\includegraphics[width=5in]{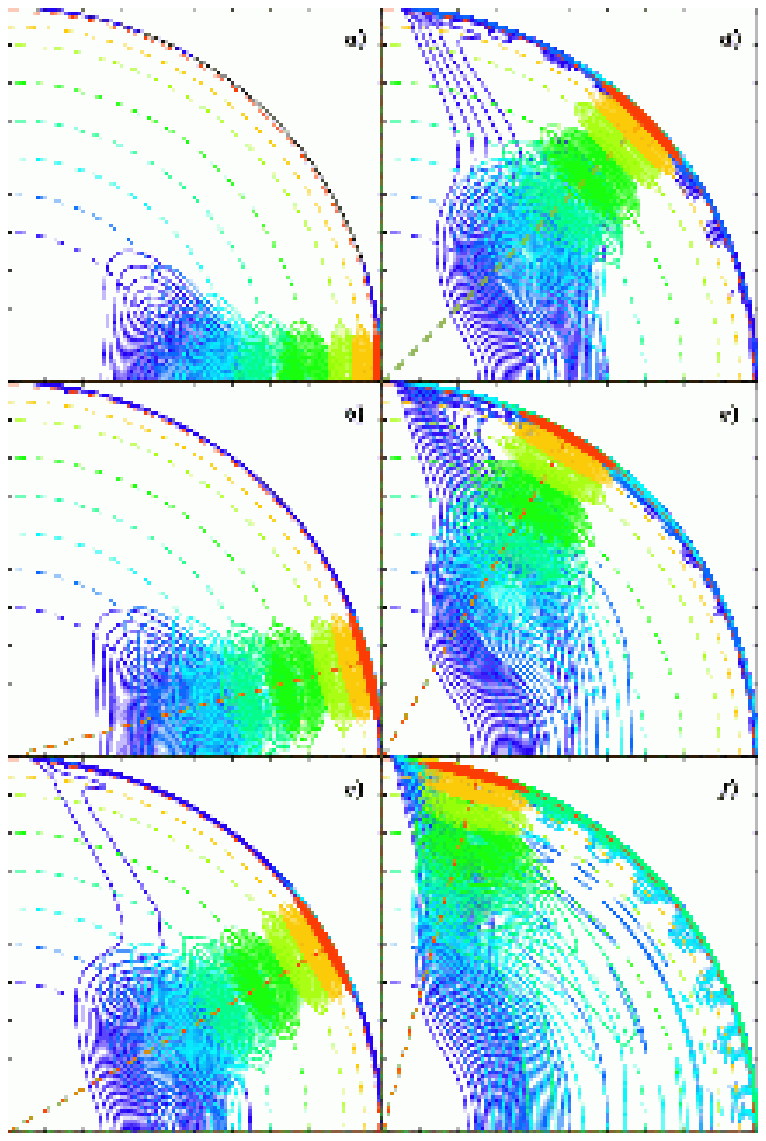}
\caption{As Figure~\ref{fig:ker1}, for a SOLA inversion.}
\label{fig:ker2}
\end{figure}

\subsection{Optimally localized averaging}
\label{section:ola}

In the Subtractive OLA (SOLA) approach \citep{1968GeoJ...16..169B,1970RSPTA.266..123B},
the minimization is applied to the difference 
 between the actual
averaging kernels ${\cal K}$ and a target kernel ${\cal T}$, for
example a 2-dimensional Gaussian or Lorentzian function. In this
case
\citep{1992A+A...262L..33P,1994A+A...281..231P} the function minimized
is 
\begin{equation}
\int_0^R\int_0^\pi{[{\cal T}(r_0,\theta_0; r,\theta)-
{\cal K}(r_0,\theta_0; r,\theta)]^2 r dr d\theta}\\
+ \lambda \sum_{i=1}^M{[\sigma_i c_i(r_0,\theta_0)]^2}.
\end{equation}
Both the tradeoff parameter $\lambda$ and the
radial and latitudinal resolution of the 
inversions must be chosen before running the inversion. If the
choice of target kernel is poor -- too narrow or too wide for the
quantity and quality of the data -- the reliability of the inversion
will suffer. In OLA inversions, setting target locations 
outside the regions that can be resolved using the data will 
result in averaging kernels displaced from their targets, and this
should be taken into account when interpreting the results.
Figure~\ref{fig:ker2} illustrates typical averaging kernels
for a 2dSOLA inversion of an MDI data set.

Another approach, older, and  more computationally expensive,
is the Multiplicative OLA (MOLA) described by \citet{1992A+A...262L..33P,1994A+A...281..231P}. Here, no target form is imposed on the averaging kernel, but it is multiplied by a term which penalizes large values away from the target location.

\subsection{Other methods}

Alternatives to full 2-dimensional inversions are the 
so-called ``1.5-dimensional'' approach, in which 1-dimensional radial 
inversions are
carried out separately for each of the coefficients describing the
latitudinal rotation variation, and ``$1d\otimes1d$'' inversions
in which the radial and latitudinal variations of the
rotation rate are integrated separately.
For details of many of these methods, please see \citet{1998ApJ...505..390S} 
and references therein.


\subsection{Limitations}
\label{section:invlimits}

It is important to bear in mind the limitations of the inversion
process when considering the results. The deepest and shallowest 
depths that can be resolved, for example, are limited by the deepest
and shallowest turning-point radii of the available modes. 
The rotational splitting at a given $m$ is to first order proportional
to the rotation rate multiplied by $|m|$;  
since the only mode whose latitudinal kernel reaches the pole is the $m=0$
mode, which has no longitudinal structure and so can convey no rotational 
information, and the modes of small $|m|/l$ 
have only a few nodes around the equator and hence have low sensitivity to the rotation, 
the 2d inversion becomes progressively less reliable
at high latitudes. Furthermore, since only modes of relatively low degree 
($l \leq 20$) penetrate into the radiative interior, the latitudinal
resolution in this region is quite poor and becomes progressively 
worse with depth; radial resolution also becomes coarser in the interior.
The practical effects of such limitations can be assessed by careful
inspection of the averaging kernels, or by performing
forward-calculation tests in which the averaging kernels are convolved
with known test profiles.

Another point to bear in mind when considering inversion results is that
the inversion can measure only the north\,--\,south symmetric part of the profile;
any asymmetry between the hemispheres is averaged out. The inversions are
also insensitive to meridional motions. Some information 
on hemispheric differences can be obtained using the the techniques of
local helioseismology, as reviewed by \citet{lrsp-2005-6}, but these
techniques, using high-degree modes, are mostly sensitive only to the
outer layers of the Sun.

\newpage


\section{Observations: A Brief Historical Overview}
\label{section:obs}

\begin{figure}[htbp]
\centerline{\includegraphics[scale=0.8]{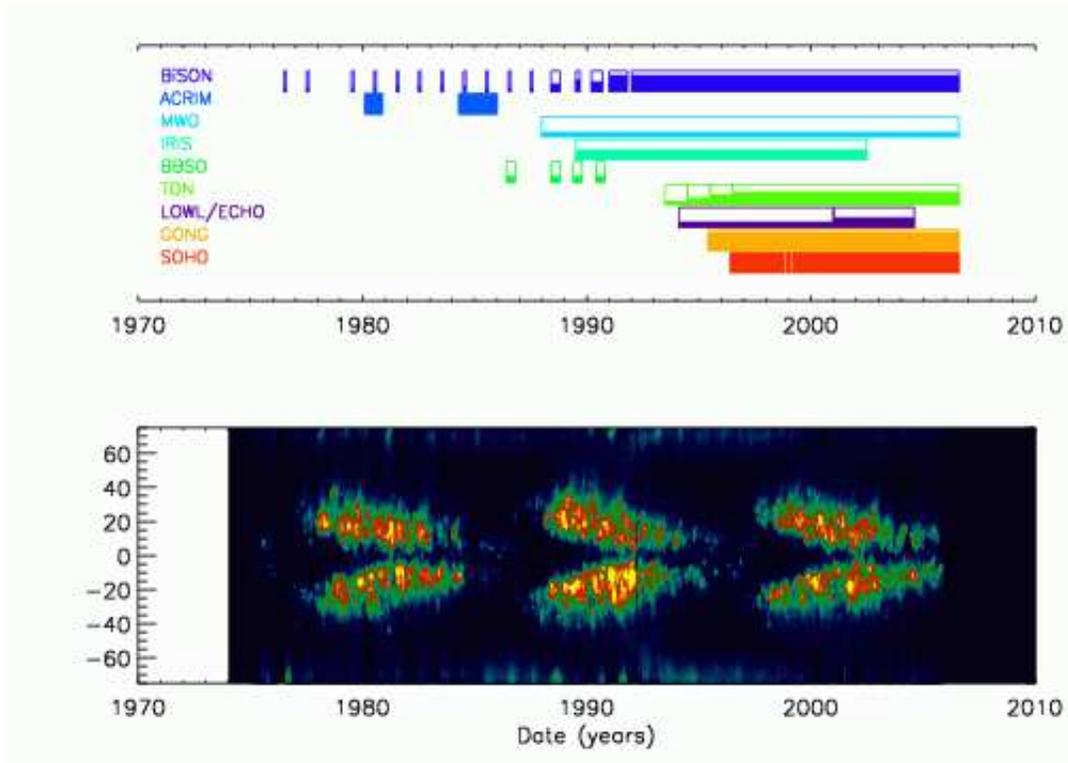}}
\caption{Schematic time line of helioseismic observations
in the last three solar cycles (top panel), with the
filled part of each bar representing approximate
duty cycle, plotted on the same 
temporal scale as the butterfly diagram (bottom panel) of the
gross magnetic field strength from Kitt Peak observations.}
\label{fig:fig1}
\end{figure}

Systematic helioseismic observations stretch back nearly thirty years,
as illustrated in the schematic chart in Figure~\ref{fig:fig1}. 

Prior to the identification of global low-degree modes by
\citet{1979Natur.282..591C}, observing runs were usually short and
carried out at a single site. However, the advantages of more extended
observations (to obtain better frequency resolution), and 
of observations not modulated by the day-night cycle, were soon recognized.
\citet{1980Natur.288..541G} and \citet{1984Natur.310...19D,1984Natur.310...22D,1986Natur.321..500D} carried out important 
observations at the South Pole during the Austral summer, but for
long time series it is more practical to observe either from
a network of sites spaced around the world, or from space. 

Some of the first long-term sets of low-degree observations 
came from the Active Cavity Irradiance Monitor (ACRIM) experiment \citep{1985Natur.318..449W,1988IAUS..123..197W} aboard the Solar Maximum Mission spacecraft, which took helioseismic measurements in 1980 and 1984\,--\,1985, 
the Mark~I instrument in Tenerife \citep{1989A+A...224..253P}, and the precursors of the Birmingham Integrated Solar Network (BiSON) \citep{1990Natur.345..322E}. Meanwhile, resolved-Sun observations were carried out at the
South Pole by Duvall and collaborators, and by various other observers
in the USA; these observations will be discussed in more detail later.

\citet{1990Natur.345..779L} observed the medium-degree modes from Big Bear Solar Observatory (BBSO) in the 1986\,--\,1990 rising phase of solar cycle~22. The first observations from widely separated sites were carried out by 
the Birmingham/Tenerife group  in 1981 \citep{1984MmSAI..55...63C}, and by 1992 the six-station
BiSON network was complete; it has been operating ever since.  Another
network of integrated-sunlight instruments, the French-based IRIS \citep{1995ASPC...76..387F}, operated from 1989\,--\,2003.

The Global Oscillation Network Group (GONG) \citep{1996Sci...272.1284H} has been collecting continuous, high-duty-cycle observations of the medium-degree $p$ modes since 1995, using a six-station worldwide network, and the Michelson Doppler Imager (MDI)  instrument \citep{1995SoPh..162..129S} aboard the SOHO spacecraft has been in operation since 1996, so that these 
two projects have essentially complete coverage of solar cycle 23. SOHO also carries instruments dedicated to the
study of low-degree oscillations; LOI (Luminosity Oscillations Imager) \citep{1995SoPh..162..101F}, and GOLF (Global Oscillations at Low Frequencies) \citep{1995SoPh..162...61G}. This wealth of high-quality data has given us the opportunity
to study the solar interior rotation and its solar-cycle changes in more detail than ever before.

Also worth noting are the LOWL-ECHO project \citep{1993ASPC...42..469T} which made medium-degree observations from one or two sites from 1994 to 2004. 
and the high-degree Taiwanese Oscillations
Network \citep{1995SoPh..160..237C} deployed over the 1993--1996 period.

All these observations will be considered in more detail as we proceed to examine the results pertaining to the interior rotation.

\newpage


\section{The Core and Radiative Interior}
\label{section:interior}

\subsection{The oblateness controversy}
\label{section:oblateness}

Interest in the rotation of the deep solar interior predates
systematic helioseismic observation. One other possible
diagnostic of the internal rotation is provided by the solar oblateness;
because the Sun is not a solid body, both gravitational 
and rotational effects cause a very slight flattening.
The lowest-order term in this effect is related to the
quadrupole moment $J_2$; confusingly, the
next-highest term, $J_4$, is sometimes called
octopole and sometimes hexadecapole. According to \citet{1997SoPh..172...11R},
who give a useful review of attempts to measure the 
solar oblateness, for a non-rotating Sun the oblateness $\Delta r = r_{eq}-r_{pol}$ is 
given by
\begin{equation}
{{\Delta r}\over{r_0}}={3\over 2}{J_2},
\end{equation}
where $r_{eq}$ and $r_{pol}$ are the equatorial and polar radii,
respectively, and $r_0$ is the radius of the best sphere 
passing through $r_{eq}$ and $r_{pol}$. If there is an additional
$\delta r$ contribution from the surface rotation this expression becomes
\begin{equation}
{{\Delta r-\delta r}\over{r_0}}={3\over 2}{J_2}.
\end{equation}
The units of $\delta r$ and $\Delta r$ are conventionally
arc ms.

\citet{1964Natur.202..432D} noted that,
if the Sun were oblate because of fast interior rotation, the effect on its
gravitational potential might destroy the agreement between the 
predictions of General Relativity and the observations of the perihelia
of the inner planets, (specifically Mercury, though Venus could
in principle experience a smaller effect) potentially leaving room for alternate theories 
of gravitation. Dicke set out to determine the solar oblateness from
ground-based measurements -- a challenging endeavor that produced
controversial results. Models (e.g., \citealt{1966ApJ...144.1221B}) 
suggested that the
interior of the Sun could still be spinning at the rapid rate at which
it originally formed, while the exterior had been slowed down by the 
torque of the solar wind. (As will be further discussed in Section~\ref{section:tachocline}, in the absence of direct 
observations of the solar interior the picture of solar
interior dynamics was not at all clear, although the existence of something 
like what we now call the tachocline
could be inferred from theoretical arguments.)
\citet{1967PhRvL..18..313D} reported finding a solar oblateness value
of $5\times 10^{-5}$,
which would be sufficient to create an 8\% discrepancy between 
observations and the Einsteinian prediction for the precession of the 
perihelion of Mercury,
and would imply a fast-rotating core.

The results, and the inferences Dicke and collaborators drew from them,
raised a storm of controversy that may well have helped to stimulate
interest in the Sun's interior rotation profile. 
The criticisms and Dicke's responses to them would fill a lengthy 
review article by themselves; we give only a few examples here.
\citet{1967Natur.213.1077R} suggested that the result might be explained
by the solar differential rotation, an idea rejected by \citet{1967Natur.214.1294D}. \citet{1967Natur.214.1297H} concluded, on the basis of
a variety of simple models of the solar ``spin-down,'' that the
Sun should have reached a state of uniform rotation quite quickly 
after its initial formation. \citet{1967Natur.216.1280S} pointed out that the
presence of magnetic field in the solar interior might well complicate
the issue, and in an accompanying article  \citet{1967Natur.216.1283G}
suggested using a space probe in a highly eccentric orbit as a more
direct test of general relativity -- or, alternatively, that 
``more complete theoretical and observational knowledge of the
visible layers and the interior of the Sun'' was needed.

At least partly inspired by the controversy, 
\citet{1967ApJ...150..551K} studied the rotational velocities
of young solar-type stars in the Pleiades and concluded that angular momentum
was lost on a timescale of about half a billion years, but noted 
in his conclusion that ``it is wrong to conclude that the present
work in any way supports the Dicke result.'' 
\citet{1968ApJ...154.1005G} considered the stability of differentially
rotating stars and concluded that it was possible but not likely that 
a radial rotation gradient such as that required by the \citet{1967PhRvL..18..313D} result might exist.

H.\, Hill, a former colleague of Dicke who had helped build the
instrument with which the 1964 observations were made \citep{1964Natur.202..432D}, and collaborators, also attempted to measure the solar oblateness, using an instrument, SCLERA [Santa Catalina Observatory for Experimental Relativity by Astrometry], which was later to play a role in the
early days of helioseismology. This measurement, carried out in 1973,
\citep{1975ApJ...200..471H}, found a $9.6\times 10^{-6}$ value for the oblateness, much smaller than that of \citet{1967PhRvL..18..313D}; \citet{1974PhRvL..33.1497H}
also pointed out a time-varying difference between the brightness
of the solar limb and poles that might account for the anomalously
high oblateness measurement.

\citet{1981ApJ...246..985U,1981ApJ...249..831U} made an early attempt to deduce what the
$J_2$ and $J_4$ terms should be based on a simple differential rotation
profile deduced from surface measurements, obtaining
predicted values of between 1 and $1.5 \times 10^{-7}$ for
$J_2$ and between $2$ and $5\times 10^{-9}$ for $J_4$ depending 
on the size of the convective envelope.

\citet{1986ApJ...311.1025D,1987ApJ...318..451D} repeated the
1966 measurements with an improved instrument, and obtained significantly
smaller values for the oblateness, with some weak evidence for a solar-cycle
variation. \citet{1996PhRvL..76..177L} made measurements
using a balloon-based instrument and obtained values of
$1.8 \times 10^{-7}$ for $J_2$ 
and $9.8\times 10^{-7}$ for $J_4$.
By this point, however, the focus in the solar oblateness studies
had moved away from trying to infer the core rotation.
\citet{2004SoPh..222..191M} used more realistic models of the internal rotation
profile to suggest that the $J_4$ term should be particularly sensitive to the
subsurface shear. 
Recent work on determining the oblateness from the shape of the solar
limb has taken into account  
considerations of near-surface temperature or magnetic variations. 
\citet{1998Natur.392..155K,2007ApJ...660L.161E} used observations from MDI during rolls
of the SOHO spacecraft and \citet{2008Sci...322..560F} used
the RHESSI X-ray telescope. The work with SOHO revealed a temporal variation in the shape of the
solar limb, with greater apparent oblateness at solar maximum,
suggesting that hotter, brighter activity belts have greater 
apparent diameter. This poses an apparent contradiction 
to the results obtained from helioseismic inferences of the
asphericity.
Indeed, \citet{2008Sci...322..560F} suggest that all the temporally-varying,
excess oblateness found in the observations can be corrected away 
by removing an ad-hoc term related to magnetic elements in the
enhanced network.

 Meanwhile, a much more flexible tool -- helioseismology
-- had become available for probing the interior solar rotation.

\subsection{Early low-degree helioseismic results}
\label{section:earlylowdeg}

Around the early 1970s there were numerous attempts to search for
global $p$-mode oscillations, with interest at first focusing on
longer-period oscillations, the low-order, low-degree modes.
Various theoretical predictions \citep{1975A+A....45...15S, 1976ApJ...204L.147I,1976Natur.259...89C}
of the periods were available, offering the hope that global oscillations
could be used to probe the rotation and structure deep inside the Sun.
At first most of the results \citep{1977ApJ...211..281L,1977ApJ...212L..95M, 1977A+A....55..411G},  were negative, except for the 160-minute
period of \citet{1976Natur.259...87S,1976Natur.259...92B}, which was later \citep{1989ApJ...338..557E} determined to
be spurious and will not be further discussed here. 
The SCLERA group \citep{1978ApJ...223..324B,1979MNRAS.186..327H,1980MNRAS.193..381C} found a variety of longer-period fluctuations in their 
solar-diameter data, but these results were not universally accepted;
for example \citet[][see also references therein]{1981A+A....94...95F}
claimed that the SCLERA results were consistent with pure noise.

Low-degree helioseismology became a reality when the Birmingham group \citep{1979Natur.282..591C} identified oscillations in the five-minute frequency band in 
integrated sunlight as low-degree global modes, using observations
from Tenerife and Pic du Midi during the summers of 1976\,--\,1978; these initial
data were adequate only to identify the spacing 
between modes of the same $l$ and different $n$, without
resolving separate $l=0$ and $l=1$ peaks.

A French-American team \citep{1980Natur.288..541G, 1981SoPh...74...59F} obtained five days of 
continuous observations at the South Pole in the austral summer of
1979\,--\,1980, and were able to identify peaks of degree 0, 1, 2, and 3 and
even a weak $l=4$ peak by superposing sections of the acoustic spectrum with different
radial order. These modes were identified as being 
of radial order around 12\,--\,30, as 
opposed to the very low-order modes that had been sought in the low-frequency
spectrum; both the noise characteristics of the spectrum and
the low amplitude of the lower-order modes mean that the
fundamental ($l=0,n=0$) mode remains unobserved to this day, although
some low-degree modes with single-digit $n$ have been identified \citep{1996MNRAS.282L..15C}. 

Soon, the Birmingham team \citep{1981Natur.293..443C}, using 28 days of integrated-sunlight
data from the Tenerife site and an analysis that 
involved ``collapsing'' segments of the acoustic spectrum so as to average
together modes of the same degree and different radial order, 
reported finding three rotationally split components in the $l=1$ 
modes and five in $l=2$, with an average separation
of 0.75~$\mu$Hz. If correct, this would have implied a 
solar core rotation substantially faster than the
surface. \citet{1982Natur.296..130I} suggested that the excess component
peaks (when two and three would be the expected number for $l=1$ and
$l=2$ respectively) could be explained if the solar core were rotating
on an oblique axis and had a very strong magnetic field; this 
idea, which was also mooted by 
\citet{1983Natur.303..292D} to explain an oscillation of about half
the solar rotation period seen in the oblateness data
\citep{1976SoPh...47..475D}, was rebutted in some detail by
\citet{1982Natur.298..350G}.


\citet{1981SoPh...74...59F} reported that initial results from 
5 days of low-degree observations at the South Pole 
suggested quite short lifetimes, about 2 days; the $l=0$ peaks appeared narrower than those of 
$l=1$ and $l=2$. \citet{1983SoPh...82...55G} later 
identified about 80 normal modes in the
South Pole data, but did not confirm the \citet{1981Natur.293..443C} rotational splitting
result, instead reporting that the $l=1$ peak seemed too narrow
to accommodate the reported splitting. 

\citet{1982Natur.299..704C} reported a periodicity of approximately
13 days in the radial solar velocity, as measured using the
resonant-scattering technique and the potassium D-line, 
and interpreted this as an effect of the solar core rotation;
however, this effect was quickly explained away 
\citep{1983Natur.301..589D,1983Natur.302..808A,1983Natur.302..810E,
1983Natur.304..517D} as an artifact caused by the rotation 
of surface features -- sunspots and plage -- across the disk.

Meanwhile, the low-degree five-minute acoustic spectrum had also been observed
using the Active Cavity Irradiance Monitor (ACRIM) aboard the
Solar Maximum Mission (SMM) spacecraft \citep{1983SoPh...82...67W}.
\citet{1983Natur.305..589W} agreed with \citet{1981SoPh...74...59F}
in finding that the modes had lifetimes of about two days, 
too short for the rotational splitting reported by 
\citet{1981Natur.293..443C} to be real.

Later work \citep{1988ApJ...334..510L,1990MNRAS.242..135E, 1997MNRAS.288..623C} revealed that the width of the peaks -- inversely proportional
to the mode lifetimes -- was strongly dependent on frequency across
the five-minute spectrum, with lifetimes of a
few days in the middle of the five-minute band and weeks or months
at low frequencies where, unfortunately, the amplitudes
of the modes are also small.
Reliable direct measurement of the low-degree splittings would have
to wait for some years, while sufficiently long, high-quality
time series of data accumulated.

\subsection{Resolved-Sun measurements}
\label{section:resolvedcore}

In the meantime, resolved-Sun 
observations provided some information about the 
rotation in the radiative interior. 
\citet{1984Natur.310...19D} reported 
observations at Kitt Peak, from 10--26 May 1983, for 
degrees $0\leq l \leq 100$. When 
plotted as a function of degree, the results show a slow decrease 
in the rotational splitting,
from the highest degrees down to about $l=6$, with an unexplained
bump at $l=11$, followed by an increase at lower degrees up to
a value of 660~nHz for $l=1$. These data, inverted
by \citet{1984Natur.310...22D}, yielded a rotation profile with much
of the radiative interior rotating at or below the surface
rate, but with a modest increase in the interior.
 A similar pattern was found by \citet{1985Natur.317..591B}, 
using 6 days of observations from the newly-developed Fourier Tachometer,
a true 2-dimensional imaging instrument that gave access to all the
azimuthal orders for degrees between 8 and 50; however, the
coincidence of the $l=11$ bump seems to have been merely 
a coincidence of noise, as it was not reproduced in the
early observations from the Big Bear Solar Observatory \citep{1986Natur.319..753L}.

\subsection{The SCLERA modes}
\label{section:sclera}

\citet{1982PhRvL..49.1794H} derived splittings from the SCLERA low-frequency
peaks, and from those inferred a core rotating at 6 times the surface rate; 
however,
\citet{1984Natur.309..530W} used ACRIM data to place an upper limit
of 2.2 times the surface rate on the interior rotation rate,
inconsistent with these splittings. Later,  \citet{1985ApJ...290..765H}
identified low-degree rotational splittings in the five-minute band of the
SCLERA acoustic spectrum, but \citet{1986Natur.319..753L} and \citet{1989ApJ...343..526B} found that these results were inconsistent with the other evidence
and were probably the result of misidentification of the modes. Given the
complexity of the spectrum in question, whose derivation from measurements
sampled 
at a few
points on the solar limb made it difficult to separate out spectra of 
different degree, this seems a likely explanation.

\subsection{Low-degree acoustic mode splittings 1988\,--\,2002}
\label{section:lowdeg8802}

\begin{figure}
\centerline{\includegraphics{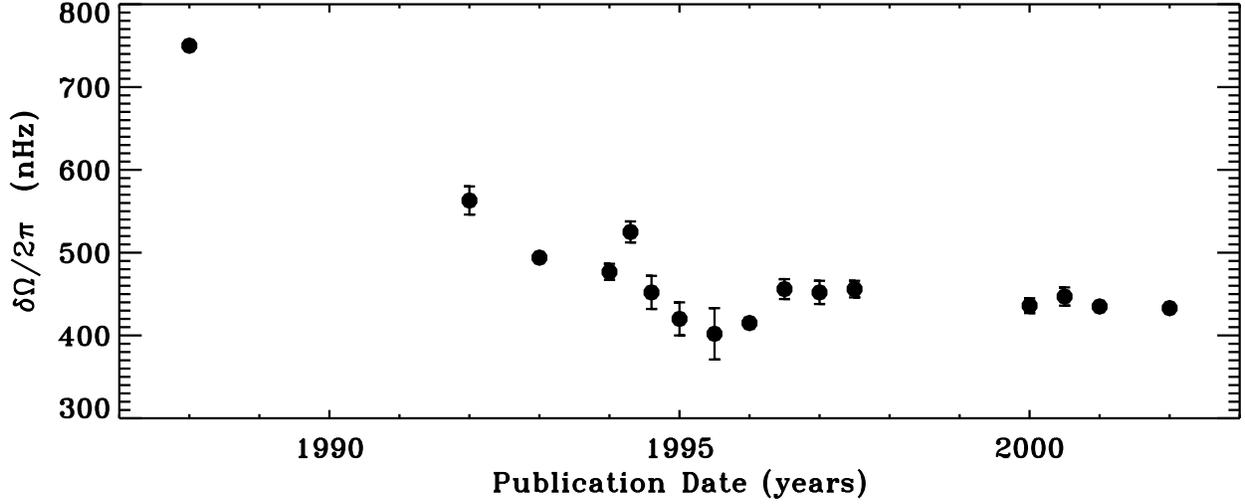}}
\caption{\label{fig:l1tab}$l=1$ splitting estimates as a function of
publication date.}
\end{figure}

The next several years were active ones for 
low-degree helioseismology, with the development of the
BiSON (Birmingham-based) and IRIS (based in Nice) networks.
Together with the IPHIR instrument that rode the
PHOBOS spacecraft on its cruise phase to Mars, and the
ground tests of the LOI (Luminosity Oscillations Imager)
instrument that would later
be mounted on the SOHO spacecraft, these brought a succession of estimates of the
low-degree splitting, as summarized in Table~\ref{tab:l1tab} and 
Figure~\ref{fig:l1tab}.
In addition to the MDI instrument for medium and high-degree
observations, the SOHO spacecraft carried both LOI and 
GOLF (Global Oscillations at Low Frequencies) specifically for observing 
low-degree modes. Even though GOLF malfunctioned and could not be
operated in its intended differential mode, instead being 
confined to making Doppler observations on one side of an absorption
line, it provided some of the best available long-term, low-degree
observations. 

\begin{table}
\caption{Summary of $l=1$ splitting measurements, 1988\,--\,2002}
\label{tab:l1tab}
\vskip 4mm
\centering
\begin{tabular}{|lll|p{1.5in}|}
\hline
{\bf Reference} & {\bf Project} & {\boldmath $\delta\nu (\mu$Hz)}  & {\bf Comment}\\
\hline
\hline
\citet{1988ESASP.286..125P} & Tenerife & 0.75  & summers of 1981-1986 \\
\citet{1992A+A...257..287T} & IPHIR & $0.563 \pm 0.017$ & Intensity measurements on PHOBOS spacecraft\\
\citet{1993A+A...275L..25L} & IRIS & 0.494 & Based on 3 low-frequency modes.  \\
\citet{1994ApJ...435..874J} &  Tenerife & $0.4768 \pm 0.0097$ & Solar maximum \\\citet{1994ApJ...435..874J} &  Tenerife & $0.525 \pm 0.0127$ &  Solar minimum \\
\citet{1994A+A...284..265T} & IPHIR  & $0.452 \pm 0.020$ &  \\
\citet{1995Natur.376..669E} & BiSON & $0.42 \pm 0.02$ & \\
\citet{1995A+A...294L..13A} & LOI & $0.402 \pm 0.031$ & $l=2$ \\

\citet{1996MNRAS.280..849C} & BiSON & $0.415 \pm 0.006$ & \\

\citet{1996SoPh..166....1L} & IRIS & $0.456 \pm 0.012$ & \\
\citet{1997SoPh..175..227L} & GOLF & $0.452 \pm 0.014$ & \\
\citet{1997A+A...317L..71G} & IRIS & $0.456 \pm 0.010$ & \\
\citet{2000ApJ...535.1066B} & GOLF & $ 0.436 \pm 0.009$ & \\
\citet{2000ApJ...535.1066B} & MDI & $0.447 \pm 0.011$ & Asymmetric profile\\
\citet{2001MNRAS.327.1127C} & BiSON & $0.435 \pm 0.0036 $ & \\
\citet{2002A+A...394..285G} & GOLF & $0.433 \pm 0.002$ & \\
\hline
\end{tabular}
\end{table}

The reported results show considerable variation, but 
apart from the early Tenerife result, which was based on 
much shorter and lower-duty-cycle observations than 
most of the others, they all cluster around the surface
rotation rate, some (particularly the IRIS results) pointing to a core rotation faster than the
surface rate and some (in particular the BiSON results) 
to one substantially below
it, perhaps as low as zero. As we approach the present time and 
the observation and analysis improve, the values tend to converge on a 
splitting quite close to that which would correspond to the surface rate.
Early in this period, there was 
room to speculate \citep[e.g.,][]{1996MNRAS.280..849C} that 
the differences reflected a 
temporal variation, but this could not explain away all the 
discrepancies.

\begin{figure}[htbp]
\centerline{\includegraphics[width=5in]{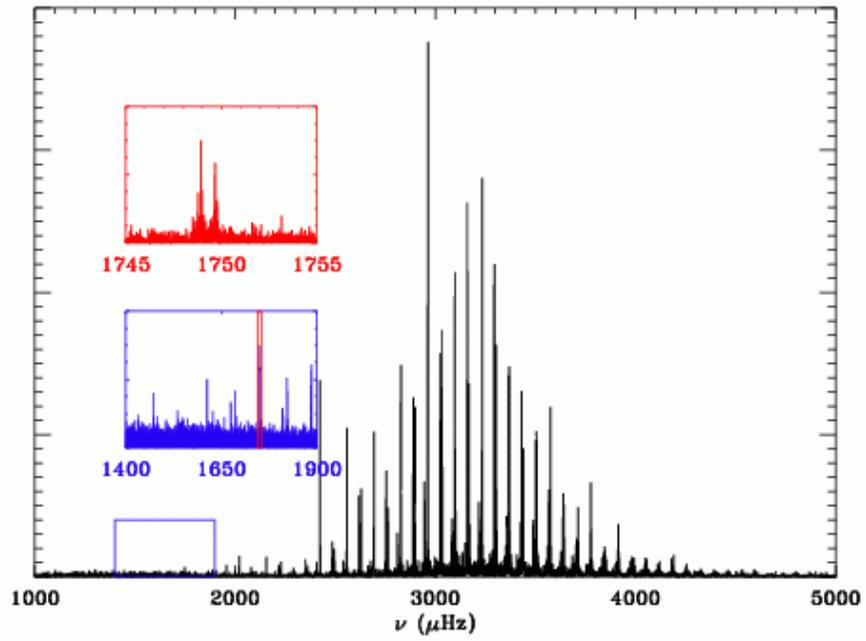}}
\caption{Power spectrum from 10 years of 
BiSON data, 1992\,--\,2002; the insets show the low-frequency end
of the five-minute band (blue) and a single, rotationally
split $l=1$ peak (red).}
\label{fig:bspec}
\end{figure}

\subsection{Pitfalls of low-degree splitting measurements}

Unfortunately, all the measurements described in 
Section~\ref{section:lowdeg8802} suffer
from similar problems, as summarized below.

\begin{enumerate}
\item \label{list1:point1}The two components of the $l=1$ mode are so close together 
(probably less than one microhertz apart) that they are resolved
only for modes below about 2.2~mHz. This has implications for the measurements:
\begin{enumerate}
\item \label{list1:point1a}Estimates of the splittings of unresolved components are
highly prone to systematic error \citep{2000MNRAS.319..365A}.
\item \label{list1:point1b} The components that can be resolved have small amplitudes (Figure~\ref{fig:bspec}) 
and therefore require both observations over extended periods and
high signal-to-noise ratios.
\item \label{list1:point1c} On the other hand, these low-frequency
modes have the advantage that they show very little frequency shift
with the solar cycle, which simplifies the analysis of long time series.
\end{enumerate}
\item \label{list1:point2}Even though the low-degree modes penetrate deep into the 
solar interior, they spend most of their time in the 
outer layers of the Sun and are not very sensitive to the
core; conversely, estimates of the core rotation are
very sensitive to small errors in the splitting measurements.
\item \label{list1:point3} In order to properly estimate the rotation profile in the
deep interior it is necessary to combine the low-degree splittings
with medium-degree ones in an inversion. However, because the
low-degree modes are so few -- a few dozen at most, compared to 
a couple of thousand  medium-degree multiplets with tens of thousands
of individual frequencies or coefficients -- the need for 
extremely precise measurements is even more pressing.
Also, combining data from different instruments with different
systematic errors may cause problems, particularly if the
observations were made at different epochs of the solar cycle.
\end{enumerate}

Point~\ref{list1:point1} above was noted by 
\citet{1993A+A...275L..25L} and \citet{1995Natur.376..669E},
and point~\ref{list1:point2} by \citet{1993A+A...275L..25L}
and \citet{1996SoPh..166....1L}, who point out that ``An accuracy of about 
30~nHz, or (1~year)$^{-1}$ on the measurement of the $l=1$ rotation splitting
does not really permit, then, to discriminate between a solar core 
rotating twice as fast as the rest or not rotating at all!''
An approach to addressing point~\ref{list1:point3} 
was made by \citet{1995ApJ...448L..57T} with the 
newly-built LOWL instrument, an imaging instrument
optimized for lower degrees. They obtained
splittings for $1 \leq l \leq 100$, and inferred a rotational
profile down to $0.2\,R_\odot$, finding a rotation rate
that barely varied with radius between $0.2\,R_\odot$ and
$0.6\,R_\odot$, apart from  a low-significance bump around $0.4\,R_\odot$.


\citet{1998ESASP.418..685E} further addressed point~\ref{list1:point3} when they combined results from several different 
instruments, including GONG, BiSON, MDI, and  GOLF. They give a nice illustration of the
tendency of higher-frequency low-degree mode splittings to be biased upward
by the mode width, a point that was further illustrated by \citet{2001MNRAS.327.1127C},
and conclude that with the then-available data it is not
possible to rule out fast rotation in the core below $0.18\,R_\odot$.

\citet{1998ApJ...496.1015C} used a genetic forward-modeling approach to
analyze the LOWL data, with results favoring a rigidly-rotating core.

\subsection{A new millennium for low-degree helioseismology}
\label{section:lowdeg2k}

Starting around the turn of the century, there was a move towards
more collaborations and comparisons between different projects in 
an effort to understand the systematic errors and 
better constrain the solar core dynamics. By this time, multi-year
observations were available from GONG and the SOHO instruments, as
well as good-quality observations from BiSON stretching back to 1991.

\citet{1999MNRAS.308..405C}  combined the LOWL higher-degree
splittings with the very precise low-frequency BiSON splittings
for the lowest-degree modes, and concluded that the data were consistent
either with rigid rotation or with a slight downturn in the rotation 
rate in the core (the latter being at best a 1-$\sigma$ result); on the other hand, \citet{1998ESASP.418..741C} 
had used a very similar analysis of GOLF and MDI data to deduce
a slight increase in the rotation rate below $0.25\,R_\odot$, but
\citet{2003ESASP.517..271G}, also using MDI and GOLF data, 
obtained rather low splitting values from a 2243-day time series 
and tentatively concluded that they could rule out a high rotation 
rate in the core.

\citet{2002ApJ...573..857E}, following on from the work of \citet{1998ESASP.418..685E}, again combined BiSON, GOLF, GONG and MDI data and found a very small
downturn in rotation in the core, while \citet{2003ApJ...597L..77C}
found a flat rotation profile down to $0.2\,R_\odot$ using 
combined GOLF, MDI and LOWL data.

\citet{2003MNRAS.346..825F} investigated the
problem of fitting the poorly-resolved higher-frequency 
low-degree mode splittings to integrated-sunlight observations such 
as those from BiSON. Using genetic fitting algorithms, they were
able to reduce, though not eliminate, the bias towards 
higher splittings for these fits. They also found, in common 
with previous work, a strong anticorrelation between the estimated
splitting value and its formal error, which would tend to 
cause overestimated splittings to be more heavily weighted in inversions.

\citet{2004SoPh..220..269G}
considered two years of ``sun-as-a-star'' observations from early in the solar
cycle, obtained from GOLF, GONG, MDI, VIRGO and BiSON, and were able to
extract not only sectoral splittings but also $a_3$ and $a_5$ coefficients
from the data, suggesting that it may be possible to infer 
differential rotation even in stars from which we will never have
resolved data.

\citet{2004MNRAS.355..535C} used artificial data to address the question
the detectability of a rotation-rate gradient in the core. They concluded that, based on the best available data from ten years of observations, 
the difference between the rotation rate at $0.1\,R_\odot$ and 
$0.35\,R_\odot$ would be detectable only if it exceeded 110~nHz.

\citet{2006MNRAS.369..985C} carried out an exhaustive 
``hare-and-hounds'' exercise, in which one participant (the ``hare'' supplies the same set of artificial data to the others, the ``hounds,'' who then apply their various fitting methods without knowing the ``true'' answer, and
compare the results.
They obtained good agreement between the different techniques for 
$l=1$, but systematic differences for the $l>1$ splittings, which are attributed 
to different assumptions about the relative heights and 
spacing of the non-sectoral ($|m| < l$) components.

\subsection{Summary of the acoustic-mode results}
\label{section:lowpsum}

To summarize, the best evidence we have so far seems to imply that the 
rotation rate between about $0.2\,R_\odot$ and the base of the convection
zone is most likely approximately constant with radius and spherically 
symmetric. It is not possible to rule out a different rotation rate
for the inner core, but there is no evidence from $p$-mode observations to support such a difference. Between about
$0.2\,R_\odot$ and the base of the tachocline, no significant departure
from rigid-body rotation has been found. As discussed by 
\citet{2002ApJ...573..857E}, for example
the available constraints  already seem to rule out
the simplest models of hydrodynamic spin-down, which would show a detectable increase in the rotation rate below $0.3\,R_\odot$. Understanding both of 
the relationship between $p$ mode splittings and the interior rotation, and
of the care needed to measure them, has greatly advanced since the
early days of helioseismology, but the rotation rate of the
innermost nuclear-burning core remains uncertain.

\subsection{Gravity modes}
\label{section:gmodes}

One possible way to improve the constraints on the core rotation would be
to use $g$ modes, or gravity waves, instead of $p$ modes.
 Because these modes have their greatest amplitude in the
solar interior, they should be much more sensitive to the
core properties. Unfortunately, they also have very small amplitudes
at the surface. The history of helioseismology is littered with 
unconfirmed reports of $g$-mode identification; see, for example, \citet{1983Natur.306..651D,1988ESASP.286..339V,1995Natur.376..139T}, and the review by \citet{1991soia.book..562H}. 
The most promising recent work has been carried out using 
long time series from the GOLF instrument aboard SOHO. 
\citet{2000ApJ...538..401A} placed an upper limit
of 10mm/s on $g$-mode amplitudes based on two years of observations,
and \citet{2002A+A...390.1119G} reduced this limit further, to 
6mm/s, using 5 years of data. Most recently, 
\citet{2007Sci...316.1591G} report finding a pattern 
of  peaks with constant spacing
in period corresponding to the model-predicted spacing 
for $l=2$ $g$ modes with $\delta l=0, \delta n=1$, and with 
a splitting that they interpret as corresponding to a core rotation
rate of 3\,--\,5 times the surface rate; however, this is still 
a preliminary result in need of confirmation.

In a related paper, \citet{2007ApJ...668..594M} point out that 
the current predictions for low-order $g$-mode frequencies are 
much more consistent than was the case a decade earlier, 
resulting in a period for the fundamental $g$-mode between 34\,--\,35 minutes. This finding does make one wonder about the usefulness of the $g$-mode observations
for discriminating among models; on the other hand, it lends somewhat 
more credence to the current identification.

\newpage


\section{The Tachocline}
\label{section:tachocline}

While the bulk of radiative interior appears to rotate almost as a solid body, 
the base of the convection zone at $0.71\,R_\odot$ coincides with a region 
of strong radial shear, above which the  
convection zone exhibits a differential rotation pattern 
that depends strongly on latitude and only weakly on depth.
This shear layer is known as the {\em tachocline}, a term introduced
to the literature by \citet{1992A+A...265..106S}, who 
attribute to D.O.~Gough the correction of the earlier term 
``tachycline'' \citep{1972poss.conf...61S}. As is evident from the
date of the latter reference, the notion of a shear layer at the bottom
of the convection zone had been present in models for some time
prior to its observational discovery, though its exact location 
was somewhat uncertain.


The existence of a layer of radial shear around the base of the 
convection zone, with 
approximately solid-body rotation below it, 
 was first demonstrated by \citet{1989ApJ...343..526B}, using the data
of \citet{1987ApJ...314L..21B}; however, the significance of their
results was quite low and they were at pains to point out that
other interpretations of the data were possible.
\citet{1989ApJ...337L..53D} used BBSO data to improve the picture of rotation at the
base of the convection zone, again finding that the 
low-latitude rotation rate increased, and the high-latitude
rate decreased, towards a common value at the base of the
convection zone.
The position of the base of the convection zone was determined by 
\citet*{1991ApJ...378..413C} using sound-speed inversions of 
helioseismic frequencies from the work of \citet{1988ApJ...324.1158D} 
and \citet{1988ApJ...324.1172L}; their value of
$0.713\,R_\odot$, confirmed by \citet{1997MNRAS.287..189B}, has been accepted ever since.

The discovery of this shear layer (as pointed out by Brown {\it et al.}) 
offered a solution to the puzzle of the
apparent absence of a radial gradient of rotation in the convection 
zone that could drive a solar dynamo, leading to speculation that 
the dynamo must operate in the tachocline
region instead of in the bulk of the convection zone. 


The tachocline lies in the region where modes of $l \approx 20$ have
their lower turning points, and the resolution of the inversions is 
quite low -- about 5\,--\,10\% of the solar radius in the radial direction.
The thickness of the shear layer is therefore likely not to be resolved
in inversions, and some ingenuity (and forward modeling) 
is required to estimate it and account for the 
effect of the finite-width averaging kernels in
smoothing out the inversion inferences. The results
of various efforts to parameterize the tachocline shape at the 
equator are summarized in Table~\ref{tab:tachtab}. They mostly 
concur in placing the centroid of the shear layer slightly below
the seismically-determined base of the convection zone, and its
thickness at around $0.05\,R_\odot$.
The largest value for the thickness, that of \citet{1996ApJ...470..621W}, 
was obtained using forward calculation and direct combination
of splitting coefficients rather than a true inversion, while the
very low value of \citet{1999A+A...344..696C} was obtained
using an inversion technique specifically designed to allow a
discontinuous step in the rotation rate at the tachocline.
The analysis of \citet{1999ApJ...516..475E} was somewhat different from the
others, in that it involved calibrating a particular model of the tachocline
against the inferred sound-speed rather than against a rotation profile.

\begin{table}
 \caption{Tachocline radius $r$ and width $\Gamma$.} 
\label{tab:tachtab}
\centering
\begin{tabular}{|llllll|}
\hline
{\bf Reference} & {\boldmath $r/{R_\odot}$} & {\boldmath $\sigma_r/R_{\odot}$} &{\boldmath $\Gamma/R_{\odot}$}
& {\boldmath $\sigma_{\Gamma}/R_{\odot}$}  & {\bf Project} \\
\hline
\hline
\citet{1996ApJ...469L..61K} & 0.692 & 0.005 & 0.09 & 0.04 & BBSO \\
\citet{1996ApJ...457..440W} & 0.68 & 0.01 & 0.12  & --  & BBSO \\
\citet{1997MNRAS.288..572B} & 0.705 & 0.0027 & 0.0480 & 0.0127 & GONG  \\
\citet{1998MNRAS.298..543A} & 0.6947 & 0.0035 & 0.033 & 0.0069 & GONG \\
\citet{1998A+A...330.1149C} & 0.695 & 0.005 & 0.05 & 0.03  & LOWL \\
\citet{1999A+A...344..696C} & 0.691 & 0.004 & 0.01 & 0.03  & LOWL \\
\citet{1999ApJ...527..445C} &  0.693 & 0.002 & 0.039 & 0.002 &  LOWL\\
\citet{1999ApJ...516..475E} & 0.697 & 0.002 & 0.019 & 0.001 & MDI \\
\citet{2003ApJ...585..553B} & 0.6916 & 0.0019 & 0.0162 & 0.0032 & MDI, GONG \\
\hline
\end{tabular}
\end{table}

\citet{1998MNRAS.298..543A} and \citet{1999A+A...344..696C} found no 
 significant evidence for a variation in the position or thickness
of the tachocline with latitude, but \citet{1999ApJ...527..445C}
found a significant prolateness, with the tachocline $(0.024\pm 0.004)\,R_\odot$ 
shallower at latitude $60^\circ$ than at the equator. \citet{2003ApJ...585..553B}
also found a slightly thicker and shallower tachocline at high latitudes,
and speculated that the tachocline location might be discontinuous at
the latitude (around $30^\circ$) where the shear vanishes and changes
sign.

\subsection{Models and the tachocline}

Even the most generous estimates for the observed tachocline thickness
are small enough to pose an interesting theoretical question: what prevents
the shear from spreading further into the radiative interior, destroying
the observed uniform rotation?
The literature on tachocline modeling is extensive, far beyond the
scope of this review. In brief,
three main candidate mechanisms 
have been proposed: turbulent flows \citep{1992A+A...265..106S}; ``fossil'' magnetic fields (e.g., \citealt{1998Natur.394..755G}); and 
gravity waves, known to observational helioseismologists 
as $g$ modes (e.g., \citealt{1997A+A...322..320Z}), 
but all these scenarios have problems when considered as the sole
mechanism. 
Recent advances in computing have made possible detailed three-dimensional
simulations to explore these issues, but these models
have not yet been able to reproduce a self-sustaining
tachocline. For a review from a 
modeler's perspective, see \citet{lrsp-2005-1}.
Also, a variety of discussions of tachocline models are
collected in the book edited by \citet{2007sota.conf.....H}.

\newpage


\section{Rotation in the Bulk of the Convection Zone}
\label{section:bulk}

The surface differential rotation, with 
the equator rotating faster than the poles,  was known from, for example, 
sunspot tracking, long before helioseismology opened up the solar
interior. 
Most models in the pre-helioseismology era
predicted or assumed 
a rotation rate constant on cylinders parallel to the axis
of rotation.
This is a consequence of the so-called Taylor-Proudman constraint,
a well-known result in fluid dynamics.

\citet{1984Natur.310...19D,1984Natur.310...22D}
observed from the South Pole, using only sectoral modes; their instrument used intensity images in a Calcium absorption line, scanning rather than 
imaging the whole Sun at once. Their main conclusion was that,
``Most of the Sun's volume rotates at a rate close to that of the
surface.''

\citet{1985Natur.317..591B}
had a different instrument, the Fourier Tachometer, which 
produced $100\times 100$ pixel velocity images.  
\citeauthor{1985Natur.317..591B}'s initial crude analysis of five days of data used cross-correlation, 
and expanded the multiplet frequencies using 
low-order polynomial fits; the results showed little sign of any depth variation
in the differential rotation.

\citet{1986Natur.321..500D}, again using data from South Pole
observations but now covering the full range of 
azimuthal orders, found values of the $a_3$ coefficient (the first-order
measure of differential rotation) 
consistent with the surface rotation and rather larger than was
consistent with the results of \citet{1985Natur.317..591B}.

\begin{figure}[htbp]
\centerline{\includegraphics[width=5in]{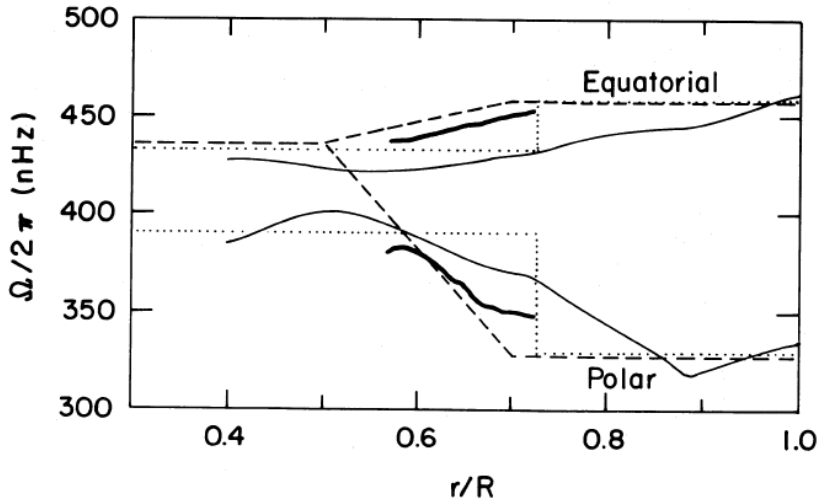}}
\caption{Summary of rotation inferences from
  \citet{1989ApJ...343..526B}, reproduced by permission of the AAS.}
\label{fig:brown89}
\end{figure}

\citet{1987ApJ...314L..21B}, with 
15  days (not all consecutive) of Fourier Tachometer data, 
could not distinguish between rotation on cylinders 
and latitude-dependence, but found that there was definitely less 
differential rotation in the radiative interior below the convection zone;
their $a_3$ values were now closer to those of \citet{1986Natur.321..500D}, and they 
declared the previous ones erroneous. 
\citet{1989ApJ...343..526B} carried out a much more detailed analysis
of the \citet{1987ApJ...314L..21B} data, 
strengthening the evidence for mostly depth-independent rotation
in the convection zone, as shown in Figure~\ref{fig:brown89}.

Both the South Pole observations and those of Brown and 
collaborators were relatively noisy and of poor resolution;
although they strongly hinted at a picture with little
radial differential rotation in the convection zone
and little differential rotation at all below it, 
other interpretations were possible.

\begin{figure}[htbp]
\centerline{\includegraphics[width=4in]{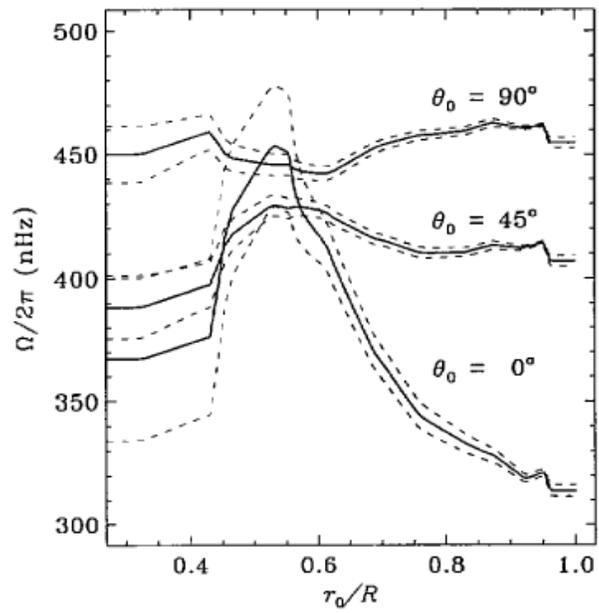}}
\caption{Rotation profile based on analysis
of BBSO splittings, \citep{1992ApJ...385L..59S}, reproduced by
permission of the AAS.}
\label{fig:scdt92}
\end{figure}

\citet{1989ApJ...336.1092L} published splittings from
 100 days of BBSO observations
in summer 1986, broadly confirming the results 
of \citet{1989ApJ...343..526B} with substantially smaller uncertainties. 
\citet{1989ApJ...337L..53D} inverted these data,
and inferred a sharp radial gradient at the base of the convection zone
and roughly constant rotation at each latitude above that. 
They also found a bump in the rotation rate in the middle of the
convection zone, to which we will return below.
Other inversions of the same data set were 
presented by \citet{1988ESASP.286..149C} and \citet{1988ESASP.286..131L},
with similar results, though not all the early inversions 
(c.f. \citealt{1988ESASP.286..117K, 1991PASJ...43..381S}) produce
such recognizable results; this may be an example of the difficulty of 
using OLA-type techniques for data with insufficient higher-degree modes.
Another (2D OLA) inversion of these data, shown in Figure~\ref{fig:scdt92}, was carried out by \citet{1992ApJ...385L..59S}, who illustrated their averaging kernels; these were rather broad,
but adequate to rule out a rotation-on-cylinders model. This 
paper was also the first to make the important point that 
the so-called ``polar'' rotation rate inferred from inversions 
is actually localized somewhat away from the pole.

\citet{1993ASPC...42..213G} continued to challenge the observers
to completely exclude rotation on cylinders, pointing out that it was
possible to construct a cylindrical model that satisfied the
constraint of the BBSO data, but \citet{1994ApJ...434..378S} showed that
such a model could not be made consistent with both the Fourier Tachometer
data and the gravitational stability of the rotating Sun. 

\citet{1993ApJ...412..870B} analyzed Fourier Tachometer observations
from 1989 and pointed out a ``wiggle''  in the 
splitting coefficients at $\nu/L \simeq 40 \mu{\rm Hz}$,
(corresponding to a turning-point radius of
about $0.85\,R_\odot$); attributed to daily modulation of the observations,
this now well-known effect accounts for the ``feature'' seen in the 
middle of the convection zone in many inversions of single-site data.

Better data, with long time series free from daily modulation,
 were obviously needed before much more progress could
be made, and with the advent of the GONG network in 1995 and the
MDI instrument aboard SOHO in 1996 such data became available.
Preliminary rotation profiles were presented by 
\citet{1996Sci...272.1300T} for GONG and by \citet{1997SoPh..170...43K} for MDI,
both showing the now familiar pattern of almost-constant rotation in the
convection zone, with shear layers both at the base of the convection
zone and below the surface.

\begin{figure}[htbp]
\centerline{\includegraphics[width=5in]{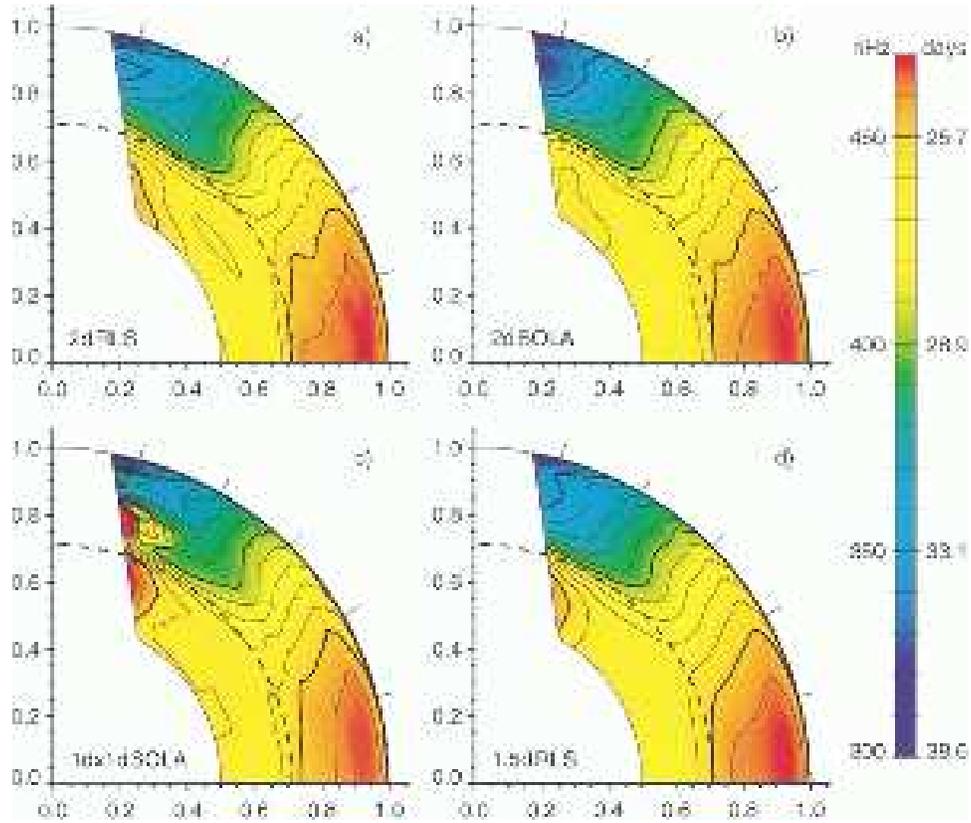}}
\caption{Four inferred rotation profiles from the first 144 days of MDI observations
\citep{1998ApJ...505..390S}; (a) 2DRLS, (b) 2DOLA, (c) 1D$\times$1D SOLA,
(d) 1.5d RLS, from \citet{1998ApJ...505..390S}, reproduced by
permission of the AAS.}
\label{fig:s98}
\end{figure}

\citet{1998ApJ...505..390S} carried out a comprehensive 
analysis of the rotation profile based on the first 144 days of
observations from MDI, using and comparing several different rotation inversion techniques with an input data set consisting of
coefficients up to $a_{36}$ for $p$ modes up to $l=194$ and 
$f$ modes up to $l=250$. 
They were able to obtain consistent and robust results from the
surface to about $0.5\,R_\odot$ at low latitudes; at higher latitudes
the domain of reliability was shallower. Roughly speaking,
the inversions could not be well localized within about $0.2\,R_\odot$
of the rotation axis.
The results (Fig.~\ref{fig:s98}) showed that the rotation in the bulk of the convection zone,
below $0.95\,R_\odot$, had a slow increase with radius at most 
latitudes, but was definitely incompatible with rotation on cylinders.

\subsection{The ``polar jet''}

In addition to the other general features described here, \citeauthor{1998ApJ...505..390S} found some evidence for a ``jet'' of faster rotation at about
$75^\circ$ latitude and $0.95\,R_\odot$; although this was more 
obvious in some inversions than in others, it did seem to have 
a signature in the coefficients themselves (see also \citealt{1998soho....6..803H}). However, this feature 
was not reproduced in inversions of GONG data \citep[e.g.,][]{2000Sci...287.2456H}, or even in inversions of MDI data analyzed with the GONG pipeline \citep{2002ApJ...567.1234S}, and it is now believed to be an artifact related to the
MDI data analysis.

\subsection{GONG/MDI comparison}

Once both GONG and MDI had been running for a few years, it became
evident that the two projects were producing inferences of the
interior rotation profile that were different in some significant 
details, particularly at high latitudes 
within the convection zone.
\citet{2002ApJ...567.1234S} carried out a careful comparison,
taking data from three epochs at different phases of the solar
cycle from each project and deriving rotational splittings or
splitting coefficients from each, both with the usual
algorithms and with those regularly used for the other project's data,
before using both RLS and OLA inversions.
The results clearly showed that most of the discrepancies arose from the
analysis pipelines rather from the data themselves. The ``CA'' peak-fitting 
algorithm used for the MDI data was able to extract modes from the
GONG data to somewhat higher degrees and lower frequencies than the
``AZ'' algorithm could manage with either GONG or MDI input data.
However, for both MDI and GONG data, the 
 ``CA'' algorithm 
introduced an anomaly in the splitting coefficients centered
at around 3.3~mHz, which in turn caused the inversion inferences
to show a higher rotation rate deep in the convection zone at higher
latitudes. Excluding these data brought the GONG and MDI data 
(analyzed with the ``AZ'' and ``CA'' pipelines respectively) into much better agreement, at the cost of somewhat degraded resolution.
Restricting both data sets to the common mode set below 3~mHz
reduced the discrepancies even farther, but did not remove the ``jet'' in the
MDI data.
Since the ``jet'' feature was only seen in the
MDI data analyzed with the CA pipeline, however, 
the authors concluded that this feature was probably spurious.

\subsection{Slanted contours}

\begin{figure}[htbp]
\centerline{\includegraphics[height=2.0in]{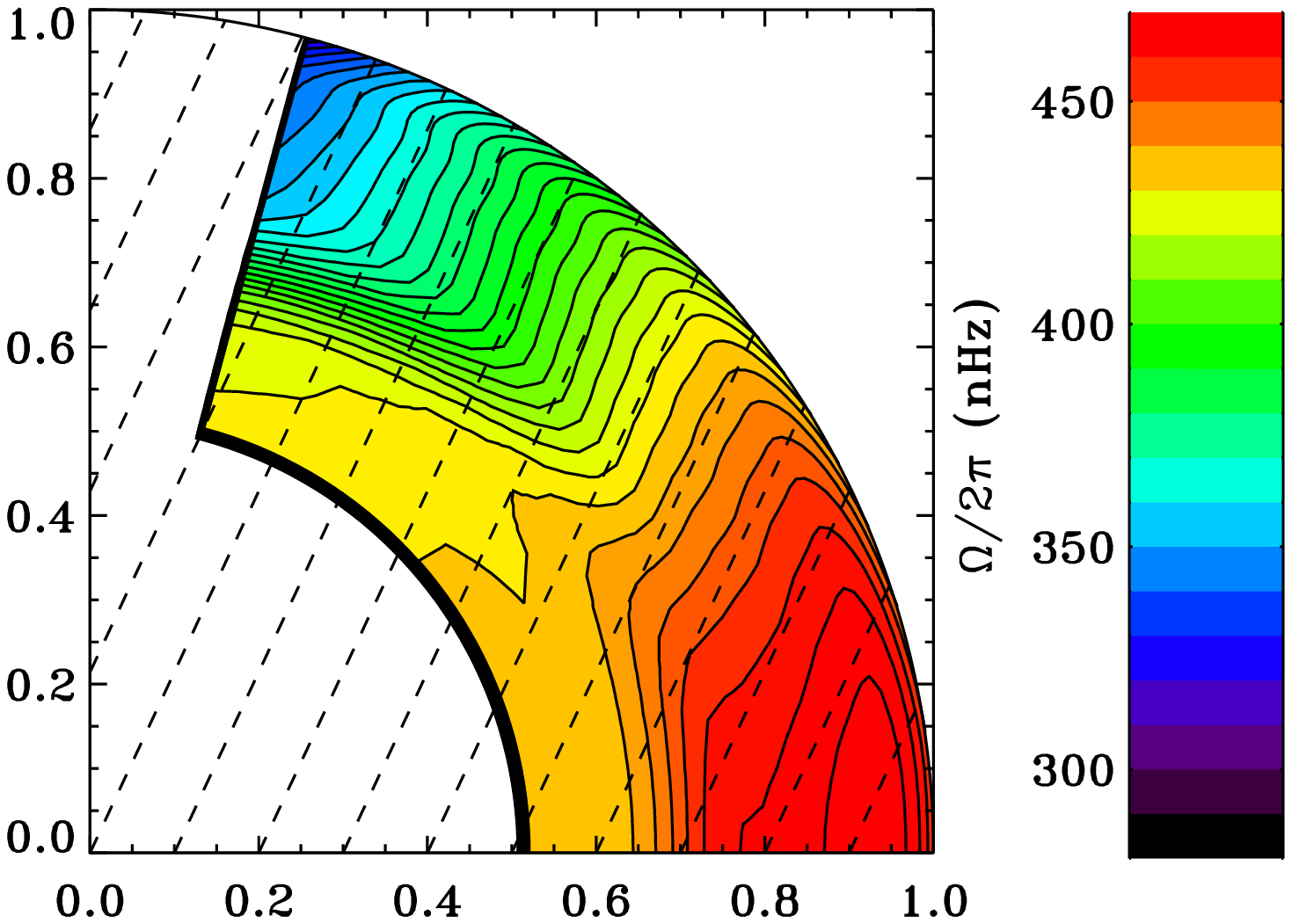}
\includegraphics[height=2.0in]{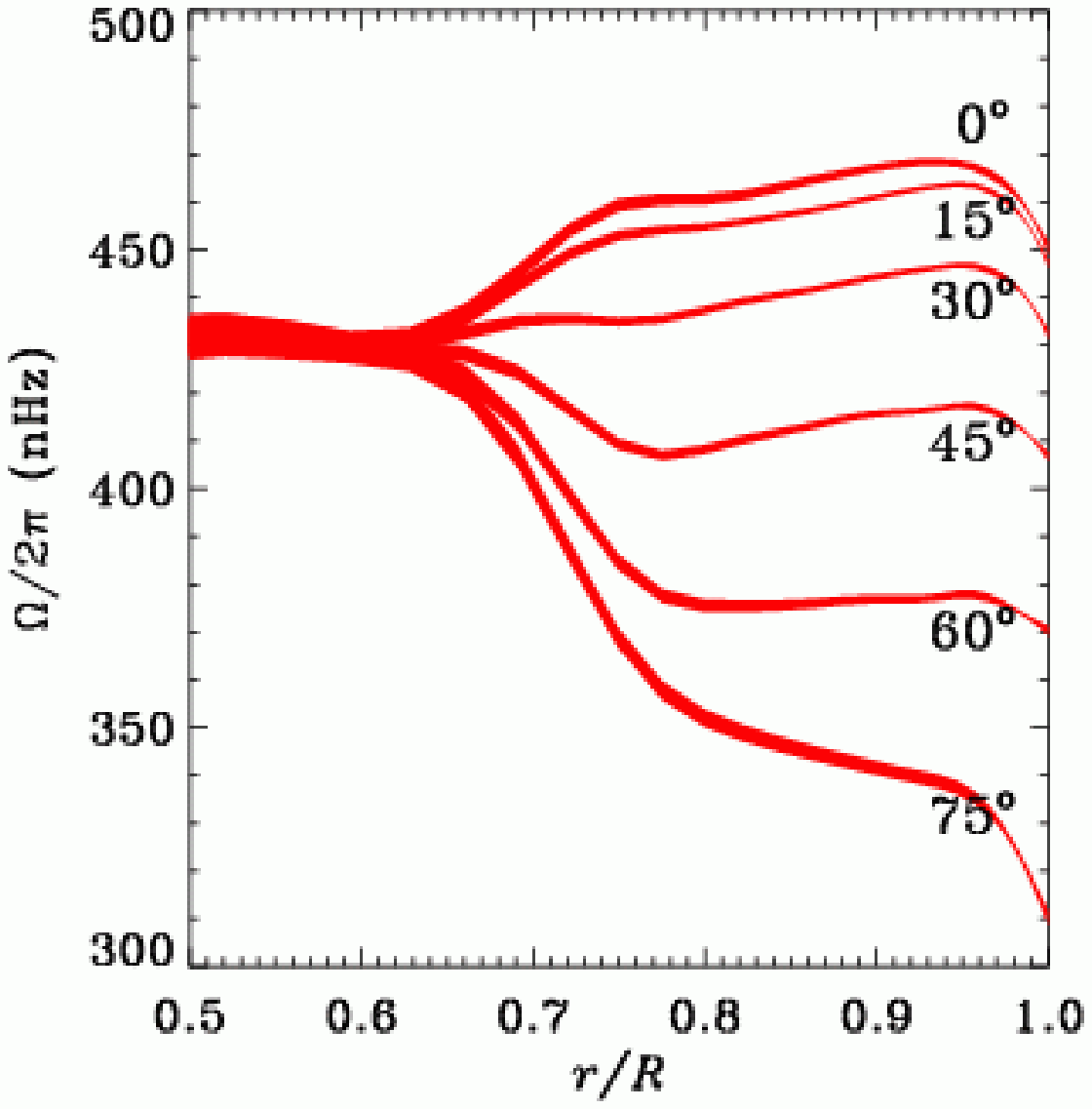}}
\caption{Mean rotation profile from GONG data;
contours of constant rotation (left),
showing lines at $25^\circ$ to the rotation axis as dashed lines,
after \citet{2005ApJ...634.1405H}, and cuts at constant latitude 
as a function of radius (right), after \citet{2000Sci...287.2456H}.}
\label{fig:slant}
\end{figure}

Although much of the debate in the early 1990s centered on 
discriminating between rotation constant on cylinders and 
rotation constant along radial lines, neither
picture gave a complete description of the data.
\citet{2003ESASP.517..283G,2005ApJ...634.1405H} pointed out that the
differential rotation in the bulk of the convection zone,
at least at low- to mid-latitudes, could be quite well described by
saying that the contours of constant rotation lay at about a
$25^\circ$ angle to the rotation axis, as illustrated in
Figure~\ref{fig:slant}.

\begin{figure}[htbp]
\centerline{\includegraphics[width=5in]{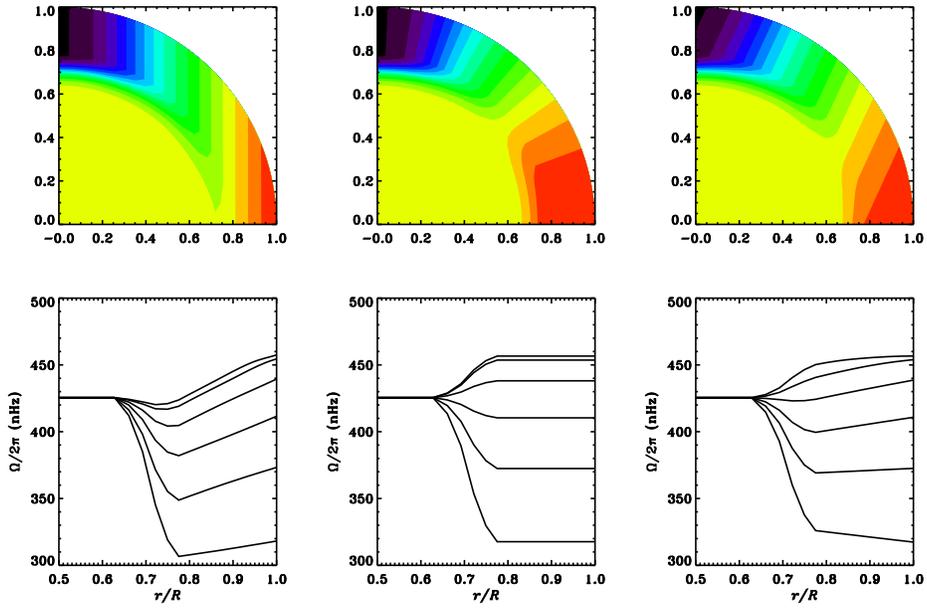}}
\caption{Idealized rotation profiles
for rotation constant on cylinders (left), radial lines (middle) and
lines at $25^\circ$ to the rotation axis (right). The top row shows
contours of constant rotation, while the lower row shows rotation rate
as a function of radius at constant latitude for latitudes at $15^\circ$ 
intervals from the equator (top) to $75^\circ$ (bottom).
The rotation rate is matched to the GONG inferences at $0.99\,R_\odot$
and smoothed to simulate the broadening effect of inversion resolution 
on the tachocline; the near-surface shear was not included.}
\label{fig:cyl}
\end{figure}

Figure~\ref{fig:cyl} compares idealized rotation profiles
for the cylindrical, radial, and slanted-contour configurations.

\subsection{Polar rotation}
\label{section:slowpole}

Another interesting feature revealed by the early GONG and MDI observations
\citep{1998ApJ...505..390S,1998ApJ...503L.187B} 
was that, while the surface rotation rate was mostly well described by 
the usual three-term expansion in the cosine of the
colatitude $\theta$, $\Omega(\theta)= A + B\cos^2\theta+C\cos^4\theta$, 
(e.g., \citealt{1984SoPh...94...13S})
the rotation rate close to 
the poles was significantly slower than that. The authors 
speculated that this might be a result of drag from the solar
wind, and that the effect might therefore disappear or become 
less marked at epochs of higher activity. In fact, though the 
inferred high-latitude rate did speed up during solar maximum 
-- as seen, for example, in \citet{2005ApJ...634.1405H} and in Figure~\ref{fig:torfig2} -- it remained at
all times lower than the extrapolation of the three-term fit.

\subsection{Models and rotation in the convection zone}

The interior rotation is only one part of the complex system that 
drives the solar cycle, but it is perhaps still the easiest part to 
measure in the solar interior; the meridional circulation 
can be directly measured only in the shallower subsurface layers,
and buried magnetic fields can at best only be inferred indirectly.  
The differential rotation in the convection zone must
arise from the interaction of convection cells
and Coriolis forces, with the meridional motions
playing an important part.

Early depictions of the solar dynamo (see, for example, 
\citealt{1974SoPh...34...11K,1975ApJ...199..761D})  required
a rotation rate increasing inward, and a meridional 
flow rising at the poles and sinking at the equator, 
in order to drive the solar cycle migration of the activity
belts in the observed sense. 
This picture, taken together with rotation on cylinders,
would have meant that the observed surface differential rotation
was a superficial phenomenon, with the dynamo operating
in the unobservable deeper layers.
At this stage, there does not seem to have been a clear distinction 
made between the direction of the meridional circulation at the surface and the
direction of migration of the magnetic activity belts during the 
solar cycle, which are of course now understood to operate in
opposite directions; the poleward meridional flow at the surface
was first measured by \citet{1979SoPh...63....3D}.

The models of \citet{1985ApJ...291..300G}
and \citet{1986ApJS...61..585G}, which 
were among the first numerical simulations
of solar rotation and the dynamo, have been cited,
for example by \citet{1992ApJ...399..294W}
as dating from ``Prior to the advent of helioseismology,''
but this is not quite correct. In fact, both these
papers refer to the Duvall and Harvey data, and 
\citet{1986ApJS...61..585G} also mentions the observations
of \citet{1985Natur.317..591B}, 
suggesting that the model results could be consistent with the helioseismic
observations if there were a layer of inward-increasing
velocity below the surface and above the domain of the
simulation. The simulations in
both cases, like their precursors over the previous
several years such as that described by \citet{1981ApJS...46..211G}, produced rotation approximately constant on cylinders
and increasing outward, which would result in a dynamo wave
propagating poleward if the dynamo were operating in the bulk of the
convection zone. 
The main message that modelers in the late 1980s seem to have taken from the 
observations was that the rotation rate was increasing outward,
in agreement with the simulations of \citet{1986ApJS...61..585G} 
but in disagreement with the $\alpha$-effect dynamo picture, which
required a rotation rate increasing inward; see
\citet{1987SoPh..110...11P} for a review representing
a theorist's perspective on the state of play at this stage.
This led \citet{1986ApJS...61..585G} to suggest (not for the first time; see also, for example, \citealt{1981ApJ...243..945G})
that the dynamo might
be operating in a thin layer at the bottom of the convection zone;
this speculation was further reinforced by the later helioseismic inferences
that clearly showed this shear layer, or tachocline (see Section~\ref{section:tachocline})
and the approximately radial 
configuration of the rotation in the convection zone.

\begin{figure}[htbp]
\centerline{\includegraphics[width=5in]{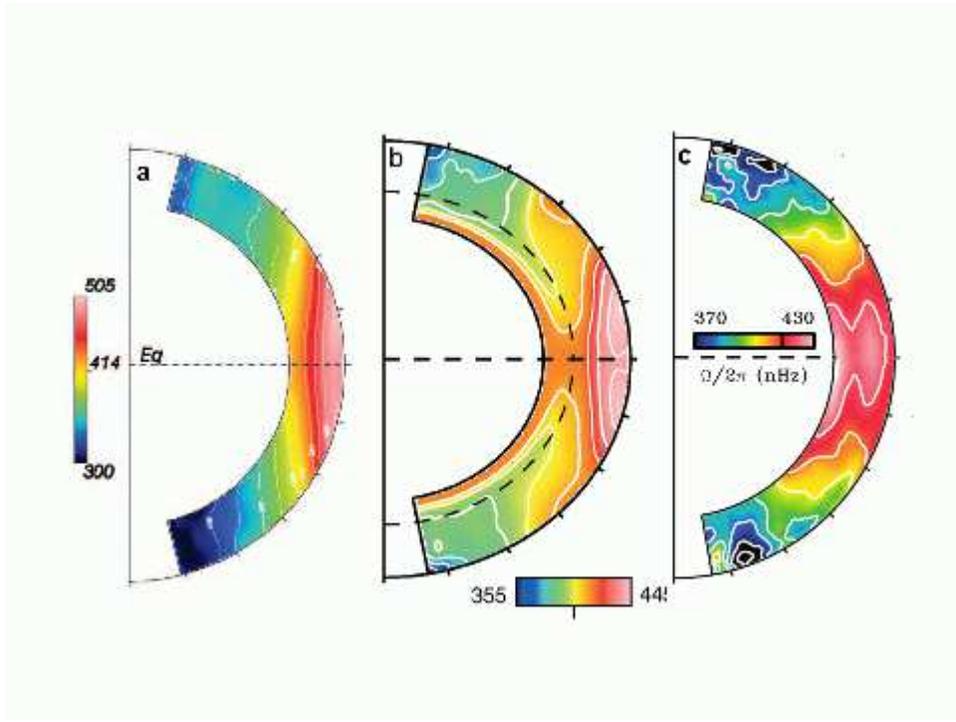}}
\caption{Three temporally averaged 
rotation profiles from the spherical-shell simulations of 
(a) \citet{2004ApJ...614.1073B}, (b) \citet{2006ApJ...648L.157B},
and (c) \citet{2008ApJ...673..557M}, reproduced by permission of the
AAS.}
\label{fig:miesch}
\end{figure}

Even quite recent global simulations of convection \citep[][for example, ]{2004ApJ...614.1073B} still show some tendency 
towards rotation on cylinders, but the higher-resolution calculation of 
\citet{2008ApJ...673..557M} mostly eliminates the cylindrical effect
and produces a rotation pattern, based on giant convection cells,  that after suitable temporal averaging looks quite solar-like, as illustrated in 
Figure~\ref{fig:miesch}.

\newpage


\section{The Near-Surface Shear}
\label{section:nearsurf}

One persistent puzzle in the measurements of rotation at the photosphere
had been that direct Doppler measurements consistently gave somewhat slower
rotation rates than the measurements made by tracing surface features.
For example, \citet{1989ApJ...343..526B} summarized the results
of \citet{1983ApJ...270..288S,1984SoPh...94...13S} as 
\begin{equation}
{{\Omega_m}\over{2\pi}}=462 -74 \mu^2 - 53\mu^4 {\rm nHz}
\end{equation}
for magnetic features and
\begin{equation}
{{\Omega_p}\over{2\pi}}=452 -49 \mu^2 - 84\mu^4 {\rm nHz}
\end{equation}
for the surface plasma, respectively, where 
$\mu$ is the sine of the latitude.
For an overview of such measurements, see \citet{2000SoPh..191...47B}. 
The usual
explanation for the discrepancy is that while the Doppler techniques 
measure the velocity at the surface, the tracers such as 
sunspots are anchored in a faster-rotating layer deeper down.
For example, \citet{1979ApJ...229.1179G} noted that the observations
implied a subsurface shear layer and 
suggested that this might arise from angular momentum conservation in 
the supergranular layer.

An extremely early attempt to measure the subsurface rotation 
was made by \citet{1979ApJ...227..629R}, when the identification of the 5-minute 
oscillations with $p$ modes was still a relatively recent 
discovery. These authors used high-degree modes, probing about the upper
20~Mm ($0.03\,R_\odot$) of the convection zone, and detected an
inwards-increasing gradient. If these measurements are reliable, they 
represent the first detection of the subsurface shear.
However, most of the early helioseismic measurements of the internal
rotation profile were restricted to a degree range that did not 
allow the near-surface shear to be resolved in inversions. 
\citet{1990ApJ...351..687R}, attempting to measure the rotation in 
the bulk of the convection zone, also saw hints of a gradient, 
opposite to that seen at the base of the convection zone, below the
surface, and 
\citet{1992ApJ...399..294W} used forward calculation techniques on the data of \citet{1987ApJ...314L..21B} and \citet{1989ApJ...336.1092L} to deduce that the rotation rate must increase inward immediately below the surface. We should remember, however, that at this time the picture of the
internal rotation profile was not as clear as it is today, and it is not always obvious 
whether interpretation of the observations as gradients of rotation
refers to the near-surface shear, the shear at the base of the convection
zone, or some unresolved amalgamation of the two. Wilson, for example,
was not arguing for a near-surface shear layer but against 
the model with rotation constant along radii. 

With the advent of GONG and MDI, measuring modes
to higher degrees than had previously been possible,
the near-surface shear could be seen in global inversions; it is
visible in the early results presented by  
\citet{1996Sci...272.1300T} for GONG and by \citet{1997SoPh..170...43K} for MDI,
in both cases apparently changing sign at higher latitudes.

\citet{1998ApJ...505..390S} found clear evidence of the near-surface
shear in inversions of MDI data. All the inversion methods agreed well
on the shear at low latitudes, but at high latitudes the
picture was complicated by the proximity of the submerged ``jet'' feature
and the methods agreed less well. The disagreement may have been 
partly due to systematic errors in the splitting coefficients.
In the comparisons of MDI and GONG data and analysis carried out  by \citet{2002ApJ...567.1234S}, the high-latitude reversal of the shear is seen only in data
analyzed with the ``CA'' pipeline; this may be partly because the ``AZ'' 
pipeline mostly fails to recover the splittings of the (narrow, 
low-amplitude) $f$-mode peaks, but the reversal persists in the MDI
data even for the restricted common mode set.

The near-surface shear (down to about 15Mm) was studied in detail by 
\citet{2002SoPh..205..211C},
using $f$ modes from MDI data. They measured the slope of the
rotation rate, close to the surface at low latitudes,  as about $-400$~nHz/$R_\odot$, decreasing to a very small value by about $30^\circ$ latitude 
and possibly reversing in sign at higher latitudes (though this 
result, seen in only the outer 5~Mm, was dependent on only the highest-degree
modes, those with $l \geq 250$).  The low-latitude rotation rate was found to 
vary almost linearly with depth in the subsurface region, while 
if angular momentum was conserved in parcels of fluid moving with 
respect to the rotation axis, it would be expected to vary 
with the inverse square of the distance from the axis. 

\begin{figure}[htbp]
\centerline{\includegraphics[width=5in]{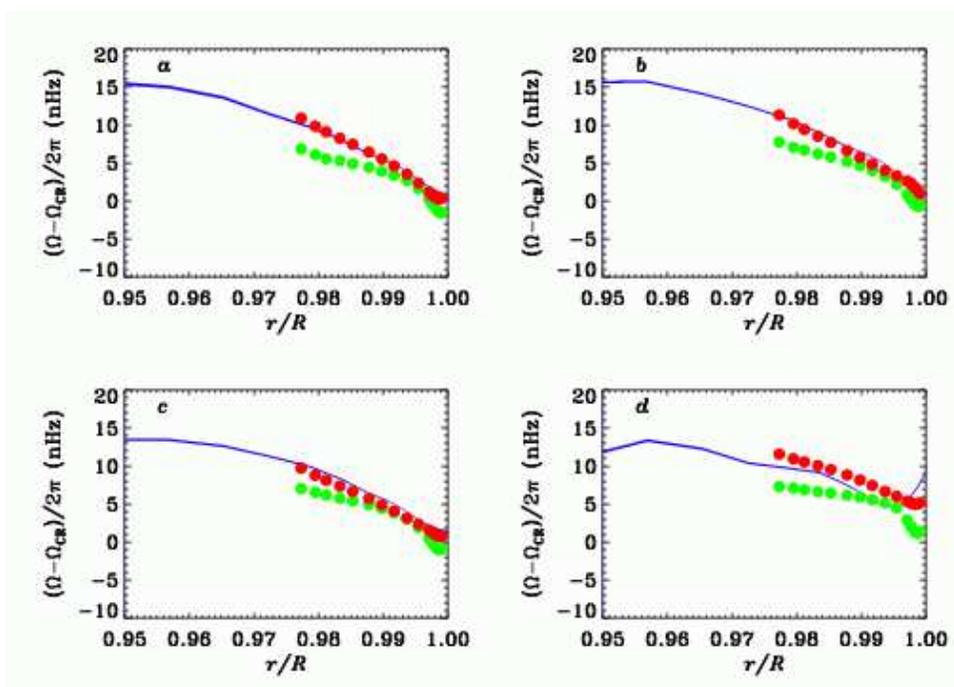}}
\caption{Radial variation of the mean rotation rate after subtraction of the 
tracking rate, for global inversions (blue) and north\,--\,south averaged local inversions 
of MDI (green) and GONG (red) data at latitudes $0^\circ$ (a), $15^\circ$ (b), $30^\circ$ (c) and 
$45^\circ$ (d); similar to \citet{2006SoPh..235....1H}.}
\label{fig:shfig}
\end{figure}

The near-surface shear is also accessible to the
methods of local helioseismology, at least for latitudes below
50--60$^\circ$. \citet{1999ApJ...512..458B} and \citet{2006SoPh..235....1H} compared results from local ring-diagram
analysis and global inversions and found, at latitudes $\leq 30^\circ$, quite good agreement between the $d\Omega/dr$ values obtained from local and global inversion results. However, although the slope from local measurements
does show some variation with latitude (Figure~\ref{fig:shfig}), it
by no means vanishes at 52.5$^\circ$, the highest latitude at
which the measurement is made. The ring-diagram results 
allow us to consider the northern and southern hemispheres separately,
but \citet{1999ApJ...512..458B} found very little difference in the 
shear between the two hemispheres.

Some attempts have been made to use the near-surface shear
to drive or at least contribute to a solar dynamo, for example by \citet{2005ApJ...625..539B}, but
\citet{2002ApJ...575L..41D} showed that any dynamo contribution from the
shear of the outer layers could only provide a fraction of the
effect needed to power the solar cycle.

\newpage


\section{The Torsional Oscillation}
\label{section:torsional}
The so-called Torsional Oscillation is a pattern of migrating
bands of faster- and slower-than-average zonal (i.e., parallel to the
equator) flow associated with the
equatorward drift of the activity belts during the solar cycle. 
It was first 
described by \citet{1980ApJ...239L..33H}, who used 
twelve years (1966\,--\,1978) of full-disk velocity observations from the 150-foot tower
at the Mount Wilson observatory and found evidence of a pattern of
flow bands migrating towards the equator; the greatest concentration
of active regions is associated with the poleward edge of the
main equatorward-moving band. 
They initially 
interpreted the high-latitude variations as
consisting of bands of faster rotation starting at the poles and 
taking a full 22-year Hale cycle to drift to the equator.
\citet{1980ApJ...239L..89S,1980ApJ...241..811S}, observing at the 
Stanford Solar Observatory, found no evidence of 
changes in the equatorial rotation rate for data from 1976\,--\,1979, but 
as this period was close to a solar minimum, 
and the resolution of the Stanford instrument was not high, this is neither surprising
nor inconsistent with the results of \citeauthor{1980ApJ...239L..33H}. 
\citet{1982SoPh...75..161L} note that \citet{1980BAAS...12Q.473S} (at a AAS meeting), 
had ``confirmed the existence of the global velocity field,'' though this is
not apparent from the latter's published abstract.

A somewhat different pattern of velocity variations is seen when 
magnetic features rather than Doppler measurements are used to determine
the surface rotation rate, as described for example by 
\citet{1993SoPh..143...19K}, who found that the pattern derived
from magnetograms lay equatorward of that from Doppler measurements,
with the slower-than-average bands coinciding with the zones of
greater magnetic flux. 
  
Mount Wilson Doppler observations since 1986, clearly showing the
pattern of migrating zonal-flow bands, were presented
by \citet{1998ESASP.418..851U,2001ApJ...560..466U}; see 
also \citet{2006SoPh..235....1H} for updated results.
The bands extend over about $10^\circ$ in latitude, and have zonal velocities
a few meters per second faster
or slower than the surrounding material, corresponding to 
excess angular velocity of less than $0.5\%$ of the overall rotation,
or a few nanohertz.

\subsection{Early helioseismic measurements}

The first hints of the signature of the migrating flow bands 
in helioseismic data can be 
seen in the BBSO data \citep{1993Sci...260.1778W}, as was pointed out by \citet{2000SoPh..192..427H},
but these measurements do not give much information on the radial
extent of the flows. \citet{1997ApJ...482L.207K} found evidence of the flows,
a few meters per second faster than the general rotation profile,
in $f$-mode measurements from early MDI data; \citet{1998ESASP.418..775G}
found a similar pattern using the time-distance technique 
of local helioseismology, while 
\citet{1998IAUS..185..141S} and \citet{1999ApJ...523L.181S}
clearly showed that these flows were migrating in a manner consistent
with the Mount Wilson Doppler observations. 
The first radially-resolved evidence
of zonal flow migration was reported by \citet{2000SoPh..192..427H} for
GONG and by \citet{2000SoPh..192..437T} for MDI, while \citet{2000ApJ...533L.163H}
combined MDI and GONG data and concluded that the equatorward-migrating part of the flow pattern (at latitudes below about $40^\circ$) penetrated to 
at least $0.92\,R_\odot$ (56~Mm below the surface). \citet{2000ApJ...541..442A} also reported similar findings.
\citet{2001ApJ...559L..67A} studied the evolution of the 
variations 
poleward of $50^\circ$, which had much higher amplitudes
than the equatorward-moving flows and which showed signs of propagating poleward over
time. The larger amplitude of the high-latitude signal may be related to the 
smaller angular momentum closer to the rotation axis.

\subsection{Recent results}

As more data accumulated, the signature of the torsional oscillation
pattern in the helioseismic observations became clearer.
\citet{2002Sci...296..101V} studied the evolution of the flows in MDI data from 1996 through 2001. They concluded
that at least the high-latitude region of changing rotation involves the whole depth of the convection
zone. The results on the radial extent of the flows at lower latitudes were less clear, with evidence
that the bands of slower rotation might penetrate close to the base of the convection zone, while
the bands of faster rotation appeared to reach about $0.9\,R_\odot$ but no deeper. Another interesting feature of that paper was the introduction of the use
of 11-year sinusoids to characterize the variation of the rotation rate
at any given location. This innovation had the useful effect of clarifying the 
pattern, making obvious the poleward propagation of the high-latitude 
flows even with data from little more than half a cycle. The existence
of a weak third-harmonic component to the eleven-year cycle, however,
was not confirmed in later work.

\begin{figure}[htbp]
\center\includegraphics[width=5in]{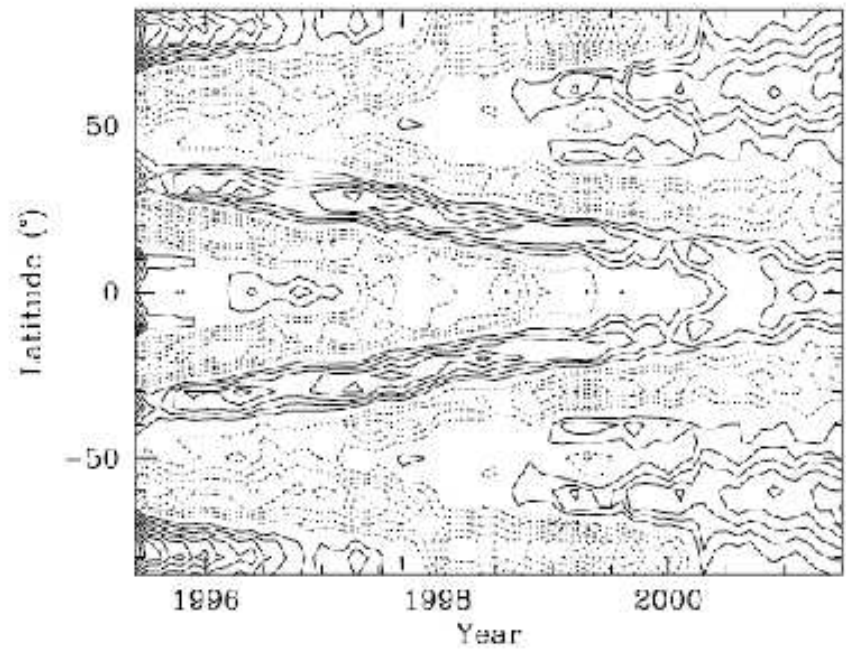}
\caption{Contour diagrams of constant rotation velocity residuals at
  $0.98\,R_\odot$, obtained using two dimensional RLS inversion of the
  GONG data, from \citet{2003ApJ...585..553B}, reproduced by
  permission of the AAS.}
\label{fig:basu03}
\end{figure}

\begin{figure}[htbp]
\center\includegraphics[width=3in, angle=90]{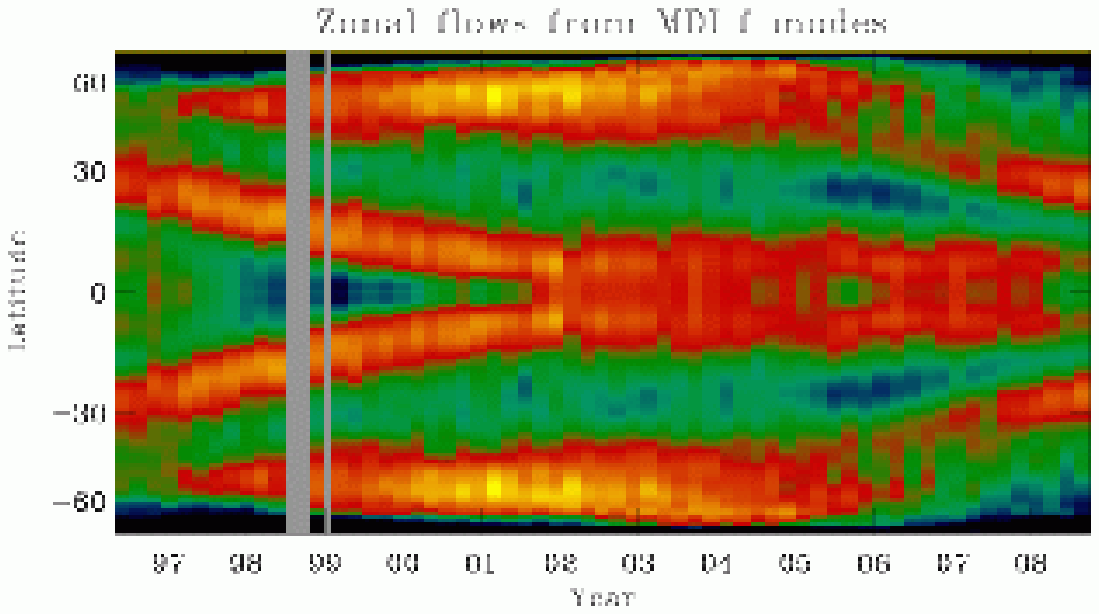}
\caption{Zonal flow pattern derived from MDI $f$-mode
measurements, with smooth profile subtracted. 
Based on a figure from \citet{1999ApJ...523L.181S}, 
updated and used by kind permission of J. Schou (2008, private
communication.)}
\label{fig:schoucolor}
\end{figure}

 \citet{2003ApJ...585..553B} found similar results in MDI and GONG data up to 2002, as seen in Figure~\ref{fig:basu03}. These 
results also hint at another subtlety; at 
low latitudes, the phase of the flow pattern is not constant along 
radial lines. In fact, the variation in the lower part of the convection 
zone appears to lead that close to the surface by a year or two, 
with the low-latitude band of faster rotation following roughly the 
same $25^\circ$ slant as the rotation contours.
This tendency was further studied by \citet{2005ApJ...634.1405H, 2006ApJ...649.1155H}, who compared inversions of MDI and GONG data 
with forward-modeled profiles based on different flow
configurations, including some derived from
dynamo models. Although some detail was lost and distorted due to the
resolution and uncertainties in the inversions, the authors were able to  
conclude that the low-latitude branch probably penetrates through 
much of the convection zone, but is sufficiently displaced in phase at
greater depths that the correlation between the surface pattern and that
deeper down almost vanishes. In this work, the 11-year sinusoid analysis
showed evidence of a second-harmonic component rather than the
third harmonic reported by \citeauthor{2002Sci...296..101V}.

\begin{figure}[htbp]
\centerline{\includegraphics[width=5.5in]{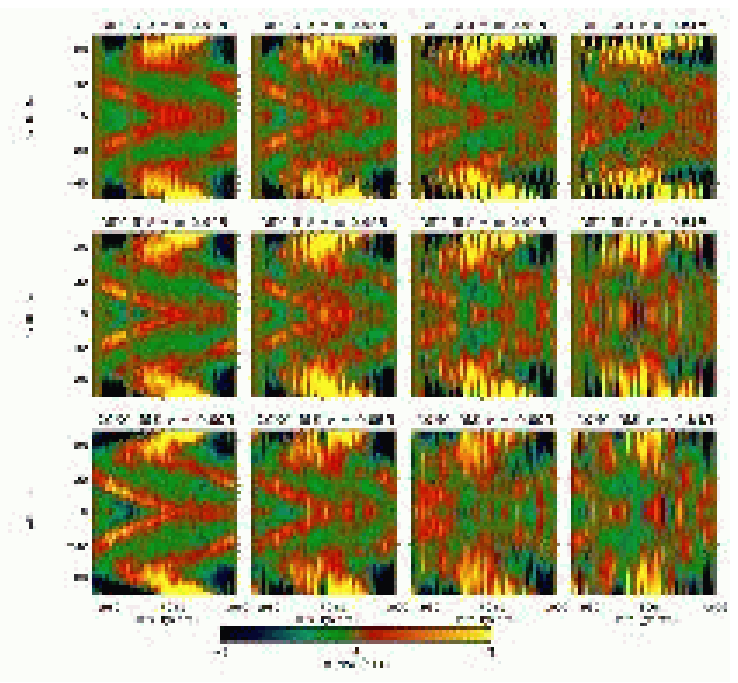}}
\caption{Rotation rate after subtraction of a temporal
mean at each location, as a function of latitude and time at 
selected depths, for OLA (top) and RLS (middle) inversions of MDI data,
and for RLS inversions of GONG data (bottom).}
\label{fig:torfig1}
\end{figure}

\begin{figure}[htbp]
\centerline{\includegraphics[width=5.5in]{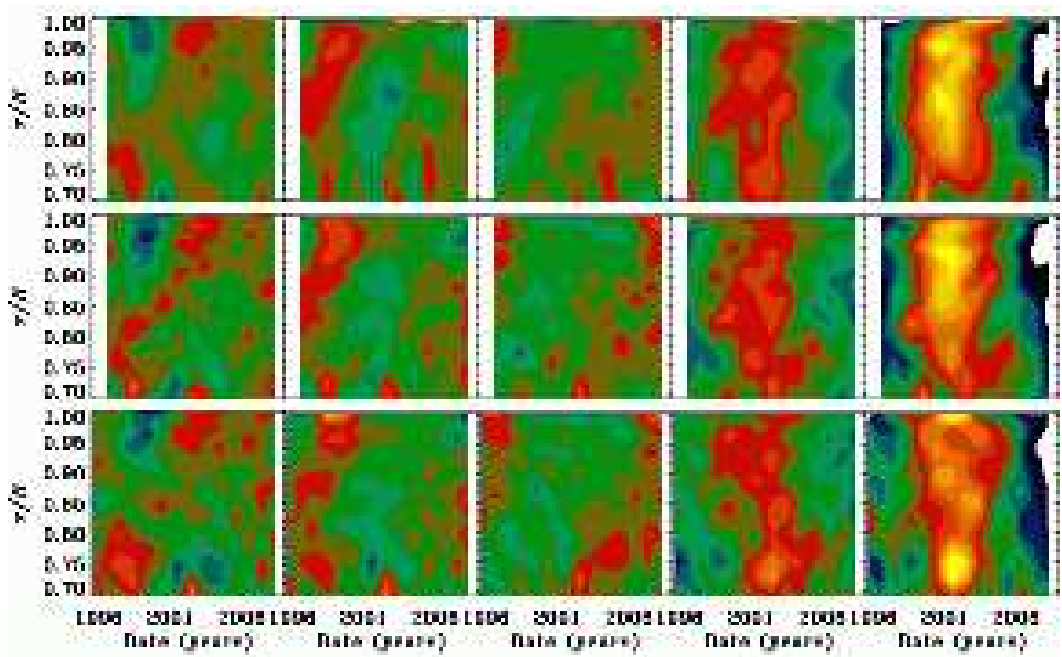}}
\caption{Rotation rate after subtraction of 
a temporal mean at each location, as a function of depth and 
time at selected latitudes. Latitudes are $0, 15, 30, 45, 60^\circ$ from left to right;
inversions are MDI OLA (top), MDI RLS (middle) and GONG RLS (bottom).}
\label{fig:torfig2}
\end{figure}

\begin{figure}[htbp]
\centerline{\includegraphics[width=5.5in]{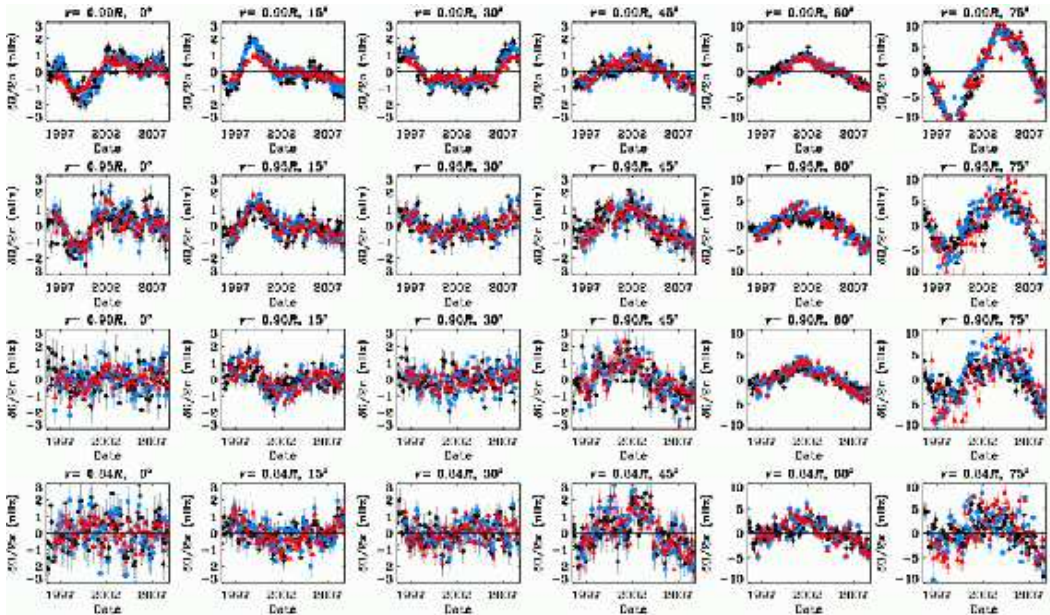}}
\caption{Rotation rates at selected latitudes and depths as a function
  of time, after subtraction of a temporal mean. The results are from
  GONG RLS (black), MDI RLS (red), and MDI OLA (blue) inversions.}
\label{fig:torfig3}
\end{figure}

\begin{figure}[htbp]
\centerline{\includegraphics[width=4in]{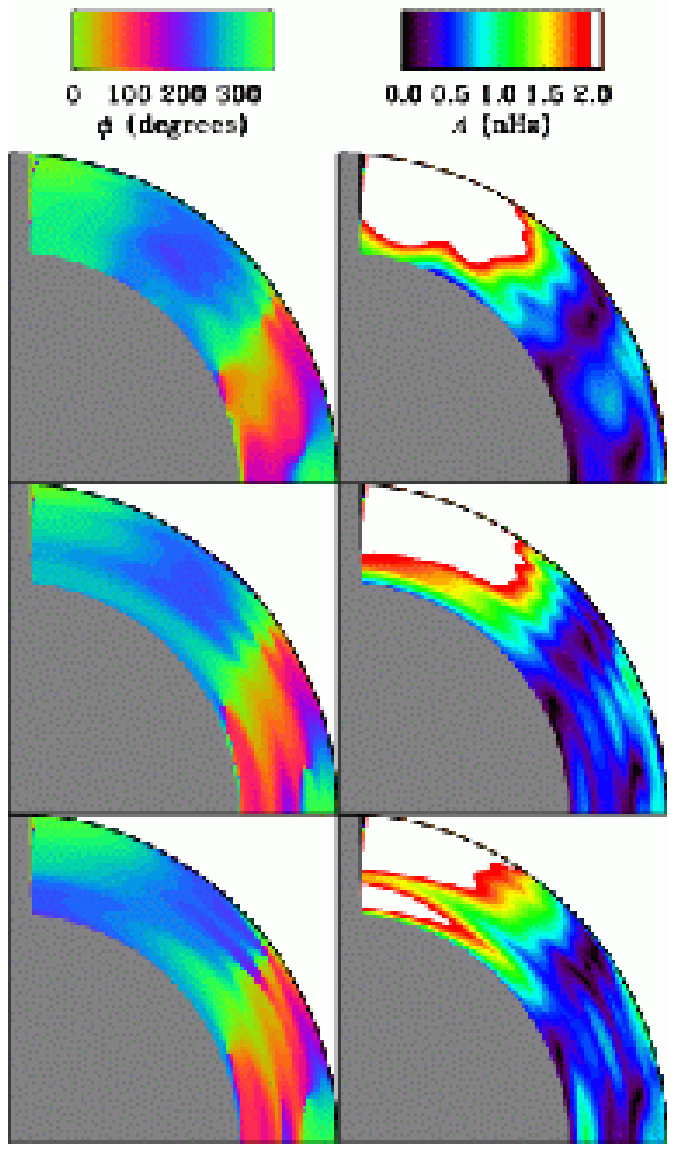}}
\caption{ Phase (left) and amplitude (right)
of 11-year sine functions fitted to temporal variation of the 
rotation rate for OLA (top) and RLS (middle) inversions of
around 11 years of MDI observations and for RLS inversions of
GONG data (bottom).}
\label{fig:phamp}
\end{figure}

Figures~\ref{fig:torfig1}, \ref{fig:torfig2}, and \ref{fig:torfig3}
show the variations in rotation rate, based on the results and figures in 
\citet{2005ApJ...634.1405H,2006ApJ...649.1155H}, but brought up to date
with the most recent GONG and MDI observations available at the 
time of writing. The plots were prepared using the same 2-D RLS inversion 
codes for both MDI and 
GONG medium-degree data, and 2-D SOLA for MDI, that were used 
for the work of \citet{2000ApJ...533L.163H} and the other related papers.
Figure~\ref{fig:phamp} shows the phase and amplitude profiles
for 11-year sine functions fitted to the rotation variations.

\subsection{Local helioseismology and the torsional oscillation}

The torsional oscillation pattern, at least at lower latitudes and
closer to the surface, is also suitable for measurements using the 
techniques of local helioseismology, in which short-wavelength, 
short-lived waves are used to infer the structure and dynamics 
of localized areas of the Sun. Because these waves do not penetrate very
far below the surface, such techniques are restricted to the
outer few megameters of the solar envelope, but this region
can be studied in much greater detail and with shorter averaging times
than is possible with global helioseismology.

\begin{figure}[htbp]
\centerline{\includegraphics[width=5in]{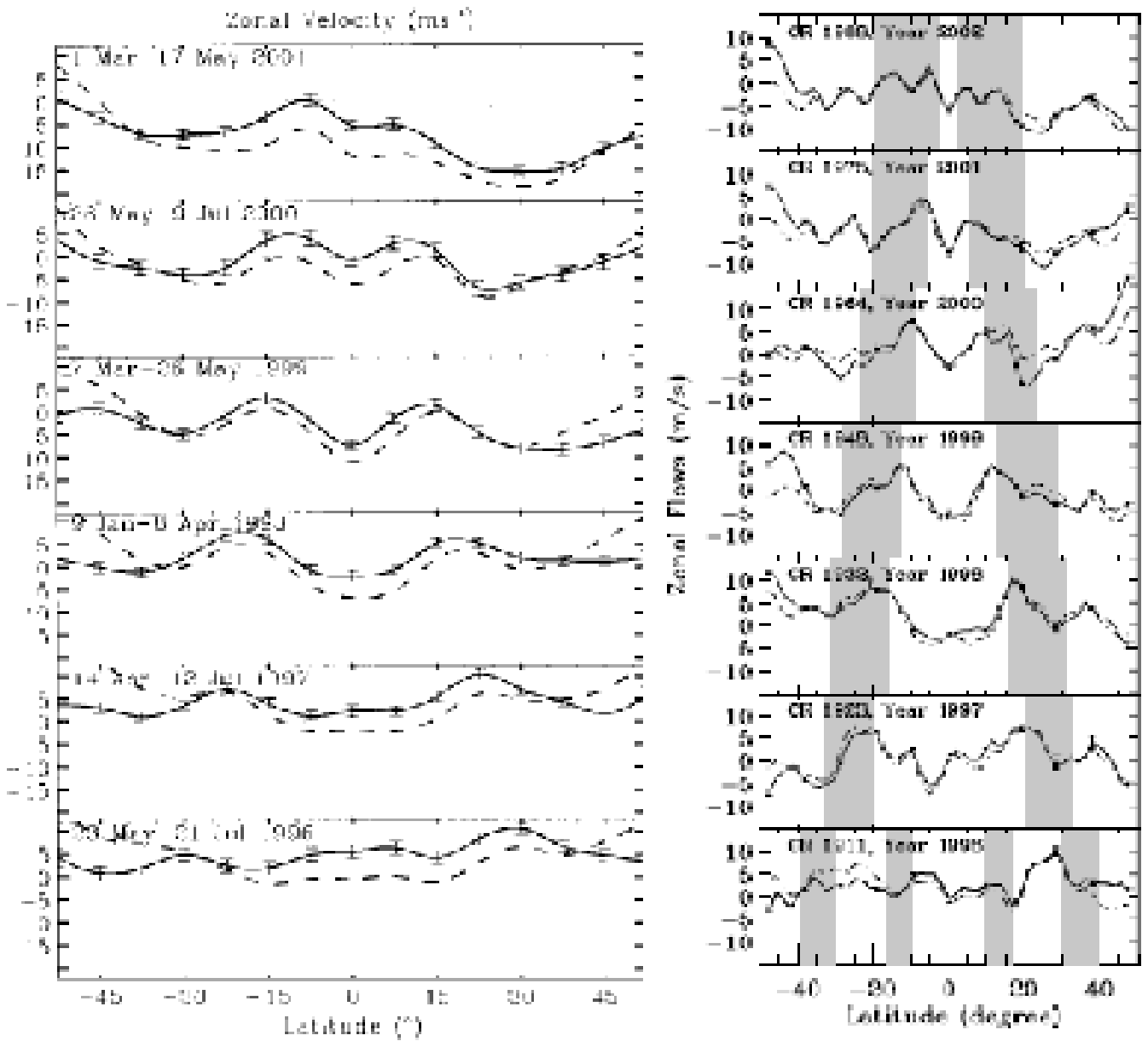}}
\caption{Local helioseismic inferences of zonal flows close to the
  surface, from  \citet{2002ApJ...570..855H} (left) and
  \citet{2004ApJ...603..776Z} (right), reproduced by permission of the
  AAS.}
\label{fig:zhfig}
\end{figure}

\begin{figure}[htbp]
\centerline{\includegraphics[width=3.7in]{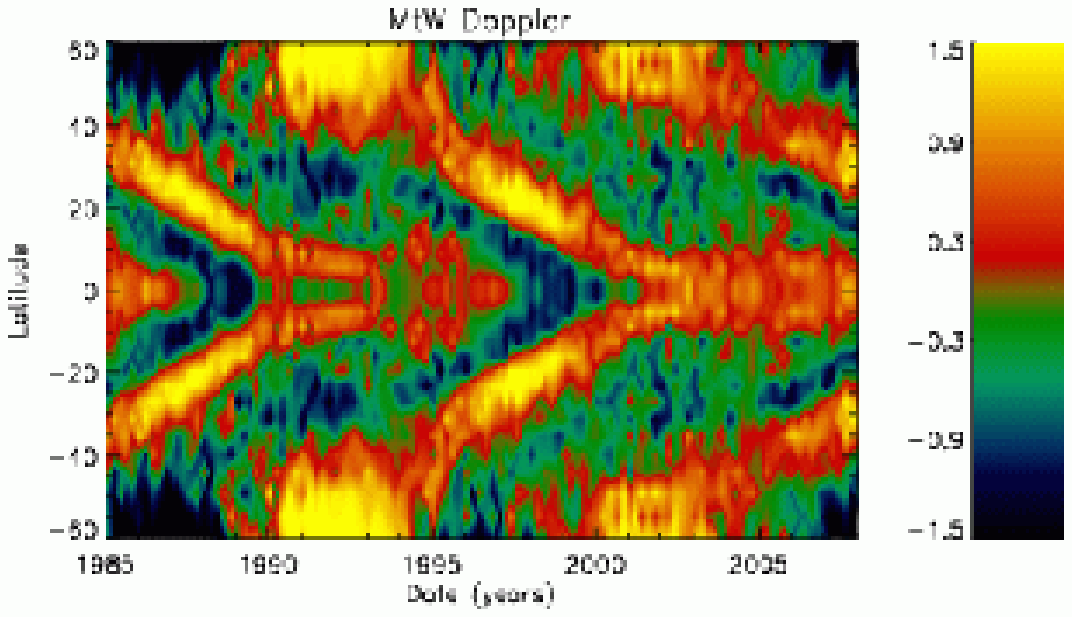}}
\centerline{\includegraphics[width=3.7in]{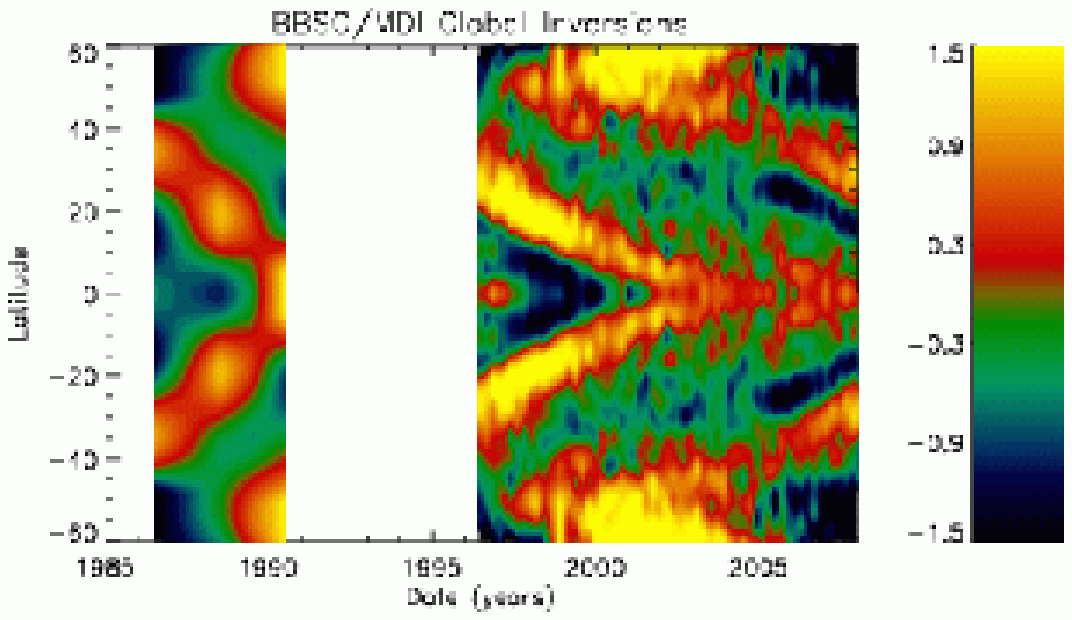}}
\centerline{\includegraphics[width=3.7in]{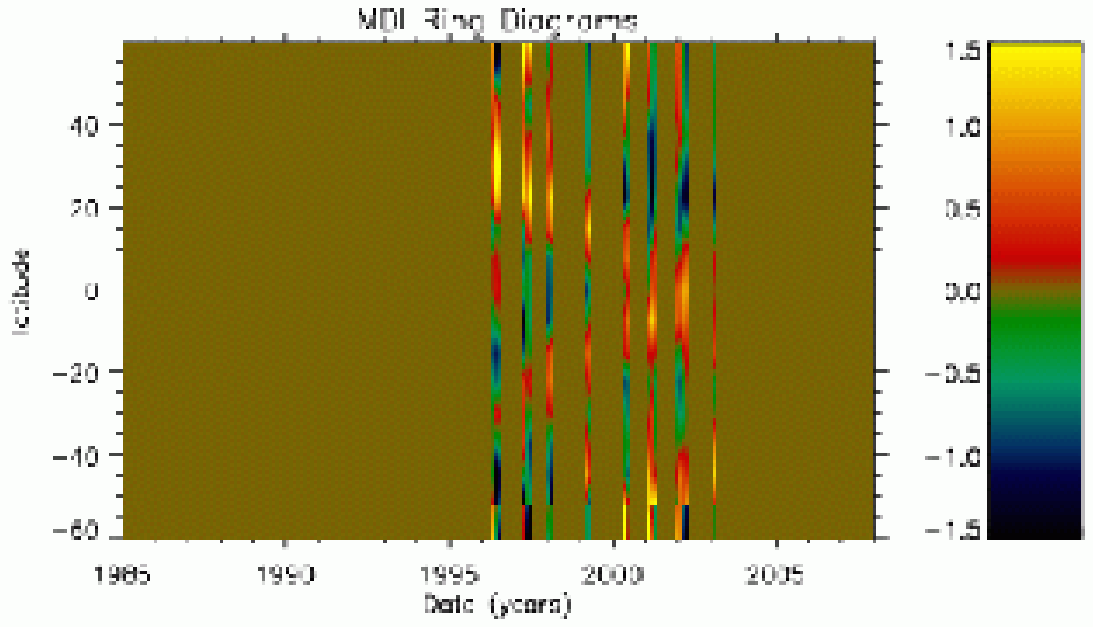}}

\caption{Zonal flows since 1986, from Mount Wilson Doppler
measurements (top), global helioseismic measurements from
BBSO and MDI (middle) and MDI ring-diagram analysis (bottom). The 
color scale is in nHz.}
\label{fig:zonefig}
\end{figure}

\citet{2000SoPh..192..469B}
detected the zonal flow migration 
using MDI data and the ring-diagram technique \citep{1988ApJ...333..996H},
in which the displacement of three-dimensional acoustic power spectra
derived from small areas of the solar disk is used to infer
horizontal flows in both the zonal and meridional directions.
Later, \citet{2002ApJ...570..855H} measured both the zonal
flows and a corresponding modulation of the meridional flow pattern,
as seen in Figure~\ref{fig:zhfig} (left). 
\citet{2002ApJ...575L..47B}, using the time-distance technique,
which considers the correlations between oscillations
at spatially separated locations, also found  bands
of  meridional flow away from the activity belts 
associated with the zonal flow bands.
\citet{2001ApJ...559L.175C,2005ApJ...630.1206C},
using data from the Taiwan Oscillations Network [TON]), 
also found diverging meridional flows associated with the activity 
belts.
\citet{2004ApJ...603..776Z} measured the zonal (Figure~\ref{fig:zhfig} right) and meridional 
flows with the time-distance technique, and reported meridional flow converging 
on the activity belts above a depth of 12Mm, with diverging flows below
18Mm, forming circulation cells around the activity belts.
The presence of inflows into the activity belts was also observed
at the surface by \citet{1993SoPh..147..207K, 1994SoPh..149..417K}.
\citet{2005ApJ...631..636K} studied the flows in about a year of
high-resolution GONG (``GONG+'') data, and concluded that the
overall flow pattern existed whether or not active regions
were included in the analysis; in other words, the zonal flow bands
and their associated converging/diverging meridional flows
appear to exist independently of the flows in the immediate vicinity of
 strong active regions.

\citet{2006SoPh..235....1H} compared the results from
ring-diagram analysis of the MDI data, global analysis of MDI and
GONG data, and the Mount Wilson Doppler observations. 
They found very similar results for the north--south symmetrized flow 
pattern close to the surface in all three observations.
Both the global and local helioseismic
data indicated that the strength of the flow pattern did not fall off
steeply below the surface.

It should be noted that the local helioseismic observations
are somewhat prone to systematic errors, some of which 
follow the changing $B_0$ angle, or tilt of the solar rotation axis
relative to the observer, as shown for example by 
\citet{2006SoPh..236..227Z}. This can result, for example, 
in a pronounced and almost certainly non-solar north--south
variation of the zonal flow measurements, which 
is generally corrected for by subtracting suitable averages.

Some further features of the torsional oscillation pattern as 
we know it from a full cycle of observations from GONG and MDI 
(and nearly two cycles of surface Doppler observations) are worth noting.

\begin{enumerate}
  \item The exact appearance of the pattern is quite sensitive to the
    background term that is subtracted. For example, compare the
    $f$-mode results shown in Figure~\ref{fig:schoucolor}, which
    were plotted as the difference from a smooth 3-term expansion of
    the rotation rate, with the plots in Figure~\ref{fig:torfig1},
    which were plotted by subtracting the temporal mean at each
    location.

  \item Although the pattern repeats -- of course not precisely --
    with each (approximately) eleven-year activity cycle, each equatorward-migrating
    flow band exists for about eighteen years, emerging at
    mid-latitudes soon after the maximum of one cycle and finally
    disappearing at the equator a couple of years after the minimum of
    the following cycle; thus, the band of faster rotation associated
    with the activity of cycle~22 was still visible at the beginning
    of GONG and MDI observations in early cycle~23, and the band that
    is expected to accompany cycle 24 became visible around 2002 (if
    we look at the mean-subtracted residuals), or 2005\,--\,2006 (if we use
    the smooth-function subtraction). On the other hand, each
    poleward-moving branch seems to last only about nine years,
    appearing a year or so after solar minimum and moving to the pole
    before the next minimum.

  \item Although the equatorward-migrating bands of faster rotation
    are clearly associated with the migrating activity belts of the
    magnetic butterfly diagram, the relationship is not completely
    straightforward. The new equatorward-propagating branch is clearly
    visible some years before noticeable new cycle active regions
    begin to erupt, and the phase/latitude profiles of the magnetic
    index and the velocity are very different. Also, as was noted by
    \citet{1982SoPh...75..161L} and by \citet{2006SoPh..235....1H},
    the strength of the torsional oscillation signal has not shown
    much change over the last few solar cycles, while the level of
    magnetic activity varies much more from one cycle to another.

  \item Although the equatorward branch of the zonal flow migration
    pattern shows some relationship to the pattern of enhanced
    activity in the Fe~{\sc xiv} corona going back to 1973
    \citep{1997SoPh..170..411A}, the ``extended solar cycle'' seen in
    these observations starts at a much higher latitude, apparently
    about $70^\circ$, before migrating to the equator over about
    eighteen years; thus even the equatorward edge of these coronal
    activity bands seems to be at higher latitude than the observed
    new branch in the zonal flows that starts at about the same time.

  \item Finally, we note that because the angular velocity changes
    associated with the torsional oscillation signal are relatively
    small compared to the difference in angular velocity between the
    surface and the bottom of the near-surface shear layer, while the
    amplitude of the signal does not decrease rapidly with depth, the
    magnitude of the shear at a given location varies by only a
    fraction of its value during the solar cycle. However, the
    fractional change in the shear is much greater than the fractional
    change in the rotation rate.
\end{enumerate}

\subsection{Models of the torsional oscillation}
\label{section:tormod}

While observers, for example \citet{1980ApJ...239L..33H} and
\citet{2001ApJ...560..466U} have speculated that the 
torsional oscillation pattern might itself be part of the
driving mechanism for the solar cycle, perhaps generating activity by 
shearing magnetic loops, modelers have generally 
seen it rather as a side-effect of the magnetic fields. 

\citet{1981A+A....94L..17S} and 
\citet{1981ApJ...247.1102Y} modeled the torsional oscillation 
as a result of the Lorentz force due to dynamo waves; 
according to the latter paper, the phenomenon would be important only
close to the surface, and would have only equatorward, not poleward,
moving bands. \citet{1982SoPh...75..161L} objected to the
\citeauthor{1981ApJ...247.1102Y} model on the grounds that it would
predict a strong correlation between the strength of the surface
magnetic field and that of the velocity signal, 
which did not seem to be the case in the observations.

\citet{1996A+A...312..615K} used a different mechanism to generate
the torsional oscillation signal in their model, considering it as the response of the Reynolds stress on the time-dependent dynamo magnetic field rather than a direct effect of the large-scale Lorentz force. This model gave a very weak
poleward branch for the torsional oscillation signal. 

Once the flows had been shown observationally to penetrate well below the surface, \citet{2000SoPh..196....1D} suggested that, ``the pattern of torsional oscillations appear to have the potential of critically discriminating between different dynamo models as, e.g., the Babcock-Leighton and interface models.''

\citet{2000A+A...360L..21C}
used a model in which the observed rotation profile
was imposed and the rotation variations arises from the action of the Lorentz force of the dynamo-generated magnetic field on the angular velocity. They were able to 
simulate approximately solar-like patterns of zonal flow bands and magnetic
activity. In subsequent papers  they focused on the the possibility
of so-called ``spatio-temporal fragmentation'' allowing cycles of different
periods in different regions, and in calculations with no density stratification
in the convection zone they found this to be feasible \citep{2001A+A...371..718C}. The effect was not too sensitive to uncertainties in the rotation law \citep{2001A+A...375..260C,2002A+A...394.1117C}, and somewhat sensitive to the boundary conditions at the outer surface \citep{2002A+A...387.1100T}. Adding density stratification \citep{2004A+A...416..775C} did not substantially change the results, though the amplitude of the oscillations in the deeper layers of the 
convection zone did decrease as the density gradient increased. However, they did find that introducing quite a small amount of
$\alpha$-quenching (magnetic feedback on turbulent convection) 
would suppress the torsional oscillation effect.


\citet{2003SoPh..213....1S} modeled the torsional oscillation pattern as 
a 
``geostrophic flow'' driven by temperature variations near the surface associated with magnetic activity, and therefore having its greatest amplitude
at the surface and falling to 1/3 of its surface value at $0.92\,R_\odot$.
This model also accounts for the observed inflows into the activity belts.
There are some problems in reconciling this model
with the observations; it is difficult to see how the
observed depth-dependent phase pattern could arise from 
a surface-originated cause, and the existence of the
flows even at epochs where there are no active regions is also hard to 
explain, though \citeauthor{2003SoPh..213....1S} suggested that the 
flows might be produced by unobserved small-scale and short-lived
magnetic regions.

\citet{2007ApJ...655..651R} used a mean-field flux-transport dynamo model, with 
a model-derived differential rotation profile and meridional flow, to investigate the effects of various driving mechanisms for the torsional oscillation. The author concluded that the poleward-propagating branch of the pattern 
could be explained by a periodic forcing at mid-latitudes without any underlying migration of buried polar field. On the other hand, in this type of model the
observed equatorward-propagating branch could not be reproduced without
adding a thermal forcing after the manner of the \citet{2003SoPh..213....1S} model. \citet{2006ApJ...649.1155H} compared such a model with the observations, and found it not to be completely consistent with the 
observed interior behavior of the flows at lower latitudes.

\newpage


\section{Tachocline Variations}
\label{section:tachvar}
\subsection{The 1.3 year signal}

\begin{figure}[htbp]
\centerline{\includegraphics[width=5in,angle=90]{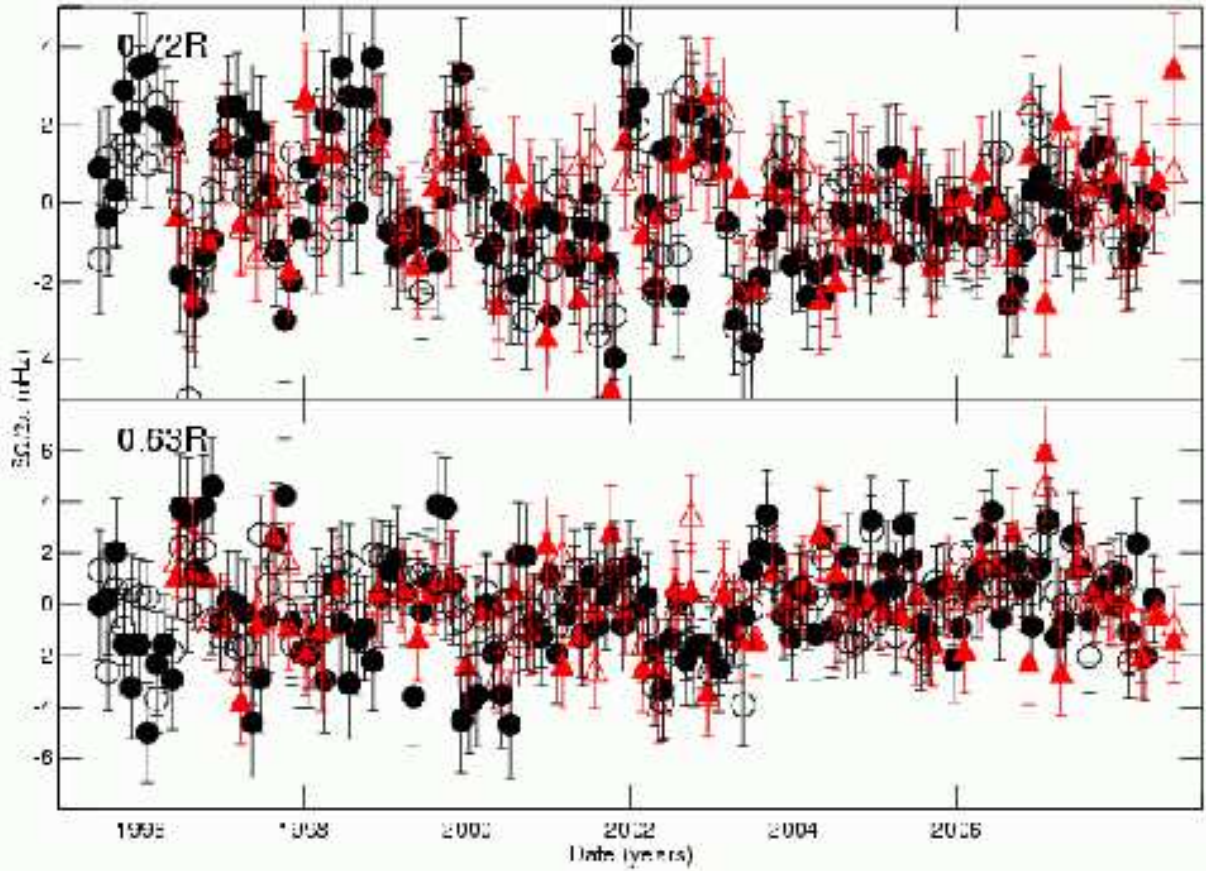}}
\caption{Rotation-rate residuals at the equator at $0.72\,R_\odot$ (top) and
$0.63\,R_\odot$ (bottom), for RLS (filled) and OLA (open) inversions
of MDI (red triangles) and GONG (black circles) data.}
\label{fig:tach1}
\end{figure}

\citet{2000Sci...287.2456H} reported finding variations of the equatorial rotation
rate close to the tachocline with a 1.3~year period during the
early years (1995\,--\,1999) of GONG and MDI observations. The strongest signal
was seen at $0.72\,R_\odot$, with a weaker anticorrelated signal below
the tachocline at $0.63\,R_\odot$. At higher latitudes, there 
was also an apparent 1-year periodicity.
The signal was more clearly seen in the GONG data, and due to the
different temporal sample of the MDI data it was difficult to make
a quantitative comparison, but the visual appearance of 
similar variations in both data sets was quite persuasive.
Figure~\ref{fig:tach1} extends the data up to the present for the
equatorial locations just above and below the tachocline.

Because of the 
role of the tachocline region in the dynamo, as 
well as the coincidence of the period with that seen in 
some heliospheric and geomagnetic observations \citep{1983JGR....88.6310S,1994GeoRL..21.1559R, 1995GeoRL..22.3001P}, this claim attracted
considerable interest, inspiring modelers such as \citet{2001A+A...371..718C}  to try to 
build models in which different periods 
could exist at the top and bottom of the convection zone. 
However, \citet{2000ApJ...541..442A} and \citet{2001MNRAS.324..498B}, with a slightly different
analysis of the same MDI and GONG data, reported finding no significant variations. (\citet{2001MNRAS.324..498B} did see a signal somewhat similar to that reported by \citet{2000Sci...287.2456H} but did not consider it significant.)

\begin{figure}[htbp]
\centerline{\includegraphics[width=5in]{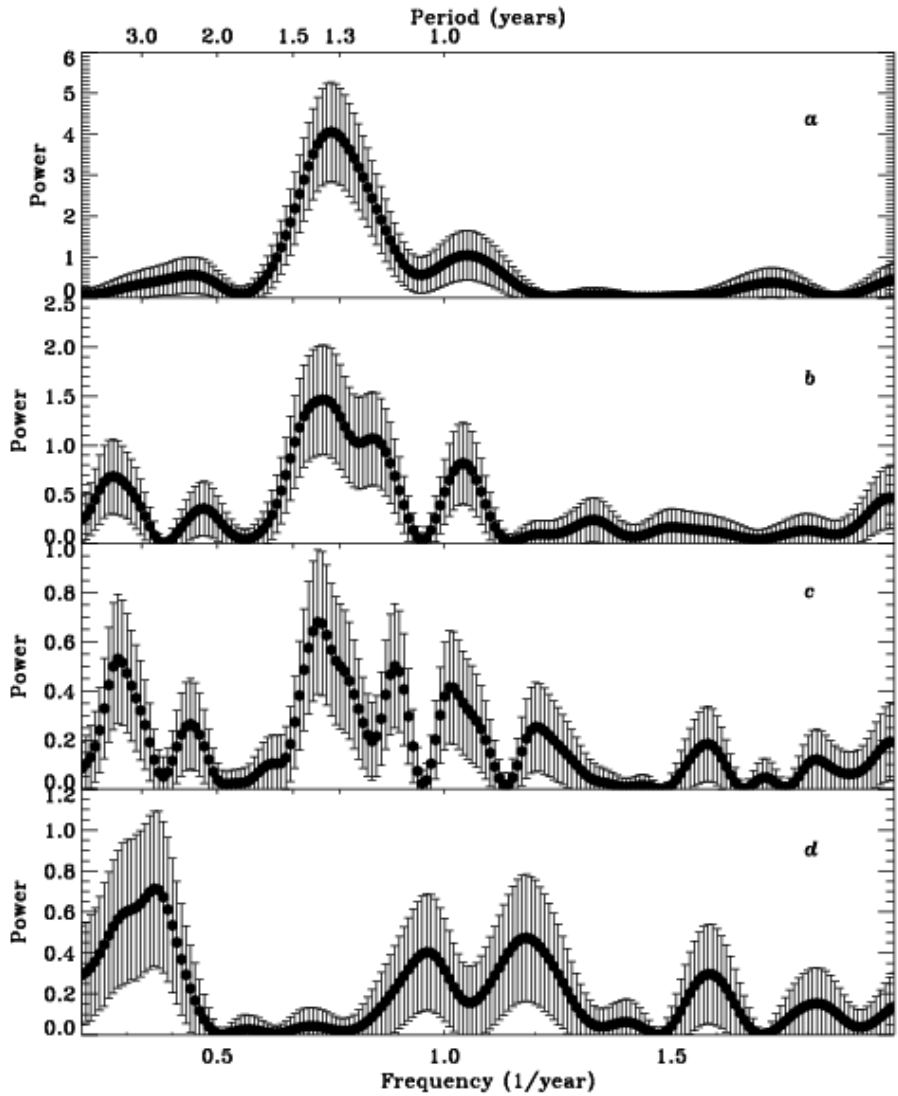}}
\caption{Sine-wave power in the rotation rate residuals from 
RLS inversions of GONG data,
at $0.72\,R_\odot, 0^\circ$, plotted as a function of frequency for 
a)~1995\,--\,2000, b)~1995\,--\,2003, c)~1995\,--\,2005, d)~2000\,--\,2005}
\label{fig:tach2}
\end{figure}

 Moreover, the periodic signal disappears in the post-2001 data  
even in the original authors' analysis \citep{2003ESASP.517..409T,2007AdSpR..40..915H}, as shown in Figure~\ref{fig:tach2}, and it seems likely that the
high-latitude 1-year period was an artifact. 
Intermittency in short-period variations is 
a known phenomenon in the geomagnetic-index data, \citep{1983JGR....88.6310S}, 
and does not
in itself imply that the phenomenon was not real. It will be interesting
to see whether the oscillation will reappear in the new solar cycle.

\subsection{Tachocline jets}

\citet{2004soho...14..376C} searched for evidence of jets close to the tachocline, which are predicted, for example, by the model of \citet{2004ApJ...610..597D}.
Using GONG data they reported finding possible evidence of a jet 
at the tachocline, migrating equator-wards by about thirty degrees in two years but not 
at the same latitude as the surface activity belts. The significance 
and meaning of this finding remain unestablished.

\subsection{Angular Momentum Variations}
Given estimates of both density and rotation as functions of depth and latitude, one can calculate the solar angular momentum locally or globally. Of course, such calculations will reflect, and in some cases enhance, any errors in the input data, and should therefore be approached with caution.

\citet{2003ApJ...586..650K} investigated the angular momentum variation based on the inversions of GONG and MDI data used by \citet{2000Sci...287.2456H,2000ApJ...533L.163H} and found variations reflecting the torsional oscillation well into the convection zone and 
1.3 year variations close to the tachocline. Because the density increases
steeply with decreasing radius, variations at greater depths will be more strongly seen in the angular momentum than in the rotation rate, but it should be remembered that no new information has been added to the data.

\citet{2007A+A...471.1011L} approached the problem from the other direction, considering the role of angular momentum transport in the modeling of the
torsional oscillation.

\citet{2008A+A...477..657A} investigated temporal variations of
the solar kinetic energy, angular momentum and higher-order gravitational
multipole moments as derived from helioseismic inferences of the
internal rotation rate; they found variations on the time scale of the solar
cycle (but not the 1.3 year cycle), with some discrepancies between MDI and GONG results. They also speculate that the kinetic-energy changes might contribute to the observed irradiance variations during the solar cycle; however, it is not clear that such a contribution is needed, as the usual view is that the
solar-cycle variation in irradiance can be modeled simply from the
effects of sunspots and plage on the surface, as discussed, for example, by 
\citet{2008SoPh..248..323J}.
 
\newpage


\section{Summary and Discussion}

Since the 1970s, helioseismology has provided several insights into 
the interior solar rotation: the approximately-rigid rotation of the
radiative interior; the differential rotation throughout the convection zone; the thin tachocline; the extension of the surface torsional oscillation
throughout the convection zone.  More than once, these discoveries have overturned
theoretical expectations, inspiring modelers to improve their
calculations in an effort to reproduce the observed behavior. Because 
of the surprising nature of many of the findings, it has been important
to have more than one source of observations, so that it is possible
to distinguish between real solar features -- especially the 
unexpected ones -- and systematic error. 

It may be that in the future solar cycle~23, with MDI and GONG operating in parallel, will be seen as a golden age of helioseismology. 
At the time of writing,
we eagerly anticipate the launch of the Solar Dynamics Observatory [SDO]
with its Helioseismic and Magnetic Imager [HMI], a successor to MDI 
that will provide
near-continuous helioseismic observations at higher resolutions than ever before and may help in unraveling the relationships between active region flows,
magnetic fields, and geoeffective solar activity as well as providing 
a continued watch on the longer-term variations in the solar velocity
fields. Sadly, however, current plans call for both GONG and MDI to cease
to collect data soon after the successful launch of SDO, which would leave HMI without any independent cross-checks, while on the low-degree front the BiSON network has recently lost its funding and there are no new dedicated low-degree space-based instruments currently scheduled. 

There are still areas -- such as the 
strength of the near-surface shear at high latitudes, the rotation
of the inner core, and any inhomogeneities and changes in the
tachocline -- that remain unclear. Furthermore, 
a complete numerical model of the solar dynamo -- vital for any
long-term predictive capability -- is still lacking, and
helioseismic observations still have an important part to play in
constraining such models as they develop.

\newpage


\section{Acknowledgements}
\label{section:acknowledgements}

This review has made use of NASA's Astrophysics Data System.

This work utilizes data obtained by the Global Oscillation Network
Group (GONG) program, managed by the National Solar Observatory, which is
operated by AURA, Inc. under a cooperative agreement with the National
Science Foundation.
The data were acquired by instruments operated by the Big Bear Solar
Observatory, High Altitude Observatory, Learmonth Solar Observatory,
Udaipur Solar Observatory, Instituto de Astrof\'{\i}sica de Canarias, and
Cerro Tololo Interamerican Observatory.
The Solar Oscillations Investigation (SOI) involving 
MDI is supported by NASA grant NAG5-13261
to Stanford University. {\it SOHO} is a mission of international cooperation
between ESA and NASA.   
NSO/Kitt Peak magnetic
data used here are produced cooperatively by
NSF/NOAO, NASA/GSFC and NOAA/SEL.

The Mt.\ Wilson observations have been supported over several decades by a
series of grants from NASA, NSF and ONR and are currently supported by
NSF/ATM.  The Mt. Wilson observatory is managed by the Mt.\ Wilson
Institute under agreement with the Observatories of the Carnegie
Institution of Washington.

BiSON has been funded by the U.K.\ Particle Physics and Astronomy Research 
Council.

\newpage



\end{document}